\documentclass[aps,pre,showpacs,twocolumn]{revtex4}
\usepackage{bm}
\usepackage{graphicx}
\bibstyle{approve.bib}
\usepackage{amssymb}
\usepackage{amsmath}
\usepackage{esint}
\usepackage{epstopdf}
\usepackage{color}
\usepackage{gensymb}
\usepackage{mathtools}
\usepackage{suffix}

\newcommand{\be}{\begin{equation}}
\newcommand{\ee}{\end{equation}}
\newcommand{\bea}{\begin{eqnarray}}
\newcommand{\eea}{\end{eqnarray}}
\newcommand{\lan}{\left\langle}
\newcommand{\ran}{\right\rangle}
\newcommand{\br}{\mathbf{r}}
\newcommand{\brc}{\mathbf{r}_{\rm c}}

\newcommand{\bk}{\mathbf{k}}

\newcommand{\e}{\varepsilon}
\newcommand{\rans}{\right\rangle_{\rm s}}

\newcommand{\tG}{\tilde{G}}

\newcommand{\pa}{\parallel}

\newcommand{\ce}{_{\rm c}}

\newcommand{\s}{_{\rm s}}

\newcommand{\hn}{\hat{\rho}}
\newcommand{\B}{_{\rm b}}

\begin{document}

\title{Nanofluidic charge transport under strong electrostatic coupling conditions}

\author{Sahin Buyukdagli}
\address{Department of Physics, Bilkent University, Ankara 06800, Turkey}

\begin{abstract}

The comprehensive depiction of the many-body effects governing nanoconfined electrolytes is an essential step for the conception of nanofluidic devices with optimized performance. By incorporating self-consistently multivalent charges into the Poisson-Boltzmann equation dressed by a background monovalent salt, we investigate the impact of strong-coupling electrostatics on the nanofluidic transport of electrolyte mixtures. We find that the experimentally observed negative streaming currents in anionic nanochannels originate from the collective effect of the ${\rm Cl}^-$ attraction by the interfacially adsorbed multivalent cations, and the no-slip layer reducing the hydrodynamic contribution of these cations to the net current. The like-charge current condition emerging from this collective mechanism is shown to be the reversal of the average potential within the no-slip zone. Applying the formalism to surface-coated membrane nanoslits located in the giant dielectric permittivity regime, we reveal a new type of streaming current activated by attractive polarization forces. Under the effect of these forces, the addition of multivalent ions into the KCl solution sets a charge separation and generates a {\it counterion current} between the neutral slit walls. The adjustability of the current characteristics solely via the valency and amount of the added multivalent ions identifies the underlying process as a promising mechanism for nanofluidic ion separation purposes.

\end{abstract}
\date{\today}
\maketitle   

\section{Introduction}

Charge transport under nanoscale forces plays a critical role in vital biological processes and various nanoscale applications. From ion exchange between cells and their surrounding aqueous medium~\cite{biomatter} to nanofluidic energy conversion~\cite{Schoch, Boc1,Boc2,Boc3,Boc4} and water desalination~\cite{Yar,Levin2006}, diverse nanoscale mechanisms involving the transport of confined liquids are regulated by the collective effect of the electrostatic, hydrodynamic, and steric forces. The accurate characterization of these nanoscale forces is thus crucial for the comprehension of anime matter and the optimization of nanotechnological approaches. This need continues to motivate intensive research work dealing with the high complexity of the nanoconfined liquids associated with the diversity of the underlying interactions. 

The minimal approach to tackle this complex transport problem consists in solving the coupled electrostatic Poisson-Boltzmann (PB) and hydrostatic Stokes equations (Eqs.).  In monovalent salt solutions governed by mean-field (MF) electrostatics, this hybrid formalism has been successfully verified by pressure-driven transport experiments~\cite{Heyden2005}, and also used to develop strategies for electrokinetic energy conversion from hydrodynamic to electrical power~\cite{Daiguji2004,Heyden2006,Heyden2007,Gillespie2012} as well as nanofluidic ion separation under hydrostatic and electrical driving forces~\cite{Gillespie2013II}.

For ions of valency $q\ce$ interacting with a membrane of surface charge density $\sigma_{\rm m}$, the importance of charge correlations responsible for the deviation from the MF electrostatics is characterized by the dimensionless electrostatic coupling parameter $\Xi=2\pi q\ce^3\ell^2_{\rm B}\sigma_{\rm m},$ where $\ell_{\rm B}\approx7$ {\AA} is the Bjerrum length~\cite{NetzSC}. In the case of silica nanochannels of characteristic wall charge density $\sigma_{\rm m}\approx1.0$ ${\rm e}/{\rm nm}^2$~\cite{ExpHey}, the coupling parameter exhibits a broad variation between the weak-coupling (WC) regime $\Xi\approx3$ of monovalent salt solutions and the strong-coupling (SC) regime $\Xi\approx195$  of quadrivalent charges such as spermine (${\rm Spm}^{4+}$) molecules. Consequently, MF electrohydrodynamic theories capable of describing monovalent charge transport fail at explaining multivalent charge-driven exotic transport behavior, such as negative streaming currents through anionic slits~\cite{ExpHey} and the electrophoretic mobility of anionic macromolecules along the applied electric field~\cite{e21,Sugimoto2019,Buyuk2018}. This defficiency indicates the necessity to use correlation-corrected theories for the rectification of the MF interaction picture.

Like-charge streaming current formation by multivalent counterion addition has been previously studied within the density functional theory (DFT) capable of accounting for charge correlations~\cite{ExpHey,Gillespie2011,Gillespie2013}. Despite the confirmed accuracy of the DFT approach, the complexity of its numerical implementation necessitates the development of alternative beyond-MF formalisms offering analytical transparency and thereby providing direct physical insight into exotic electrostatic phenomena. 

Along these lines, the field theory approach to charge liquids has been a leading alternative to DFT.  The formulation of the electrostatic field theory has been mainly motivated by the observation of seemingly counterintuitive equilibrium phenomena, such as macromolecular like-charge attraction and opposite-charge repulsion driven by ion correlations~\cite{Sim1,Besteman2005,Forsman2006}. The field-theoretic investigation of charge correlations has been initiated with the solution of the correlation-corrected PB-like Eqs.~\cite{PodWKB,attard}.  Then, a consistent one-loop (1l) theory of counterion  liquids was developed by Netz and Orland~\cite{netzcoun}. This has been subsequently extended to salt solutions symmetrically distributed around a plane~\cite{1loop} and to mixed electrolytes confined to slits and nanopores~\cite{Buyuk2012,Buyuk2014,Buyuk2015,Buyuk2018}. 

The limitation of the aforementioned theories to the electrostatic WC regime stems from the underlying perturbative inclusion of the ion correlations in terms of the coupling parameter $\Xi$. Consequently, in the SC regime of tri- and quadrivalent ions where $\Xi\gg1$, charge correlations cease to be perturbative and the loop expansion looses its validity. In order to overcome this limitation, Moreira and Netz developed a SC theory of counterion liquids derived from the expansion of the liquid grand potential in the inverse coupling parameter $\Xi^{-1}$~\cite{NetzSC}. It is noteworthy that this SC formalism can be equivalently obtained from the virial expansion of the liquid partition function in terms of the counterion fugacity. 

Biological systems are characterized by the coexistence of monovalent salt and multivalent ions. The understanding of these systems thus requires electrostatic formalisms able to take into account the multiple coupling strengths governing composite electrolytes. Motivated by this need, the SC approach of Ref.~\cite{NetzSC} has been extended by the incorporation of a monovalent salt background at the linear-MF level~\cite{Podgornik2010,Podgornik2011}. Subsequently, this dressed-ion theory has been upgraded by us via the inclusion of an additional loop correction for the monovalent salt component~\cite{Buyuk2019}. Finally, through the gaussian-level variational treatment of the WC salt, and the inclusion of the SC multivalent ions via a virial expansion, we derived a fully self-consistent formalism of complex electrolytes called the SC-dressed Schwinger-Dyson (SCSD) theory~\cite{Buyuk2020}.

In this article, we develop the SC-dressed PB (SCPB) theory corresponding to a simplified version of the SCSD formalism. Based on a restricted closure of the Schwinger-Dyson (SD) equation, this simplification enables us to account for the ionic hard-core (HC) interactions neglected in Ref.~\cite{Buyuk2020}. Within this formalism, we review the electrohydrodynamic mechanism behind the multivalent ion-driven streaming current inversion in anionic silica channels~\cite{Heyden2005,Gillespie2011,Storey2012,Gillespie2013}. We characterize the collective effect of the multivalent cations responsible for the membrane charge inversion (CI), and the no-slip constraint limiting the hydrodynamic mobility of these ions. We clarify as well the resulting electrostatic condition for the onset of current inversion, and consider the effect of repulsive polarization forces neglected in earlier studies.

The continuous demand for electronic devices of reduced size necessitates the development of polymer-based high dielectric materials for the fabrication of capacitors with small dimensions. This technical requirement has been met by coating the surface of the polymer matrices by carbon nanotubes (CNTs), enabling the enhancement of the substrate permittivity from the dielectric insulator $\e_{\rm m}\sim 2$  to the {\it giant permittivity} regime  $\e_{\rm m}\sim 10^3$~\cite{Coating}. The second part of our work probes the potential of surface-coated membrane pores for nanofludic charge transport purposes. Therein, we reveal a new current generation mechanism triggered by SC interactions. Namely, under the effect of the attractive polarization forces emerging in the giant permittivity regime, multivalent cations (anions) added to a KCl solution result in a charge separation, and generate a negative (positive) streaming current through interfacially neutral slits.

Due to the possibility to tune the sign and magnitude of this counterion current via the type and concentration of the added multivalent ions, our prediction presents itself as a useful mechanism for nanofluidic ion separation and water purification purposes. We characterize as well the impact of the same attractive polarization forces on the ionic composition of voltage-driven currents. Finally, a direct mapping from the SCPB theory to the dressed-ion formalism of Refs.~\cite{Podgornik2010,Podgornik2011} is presented in Appendix~\ref{drap}.

\section{Theory}
\subsection{Field-theoretic Coulomb model}

\begin{figure}
\includegraphics[width=1.0\linewidth]{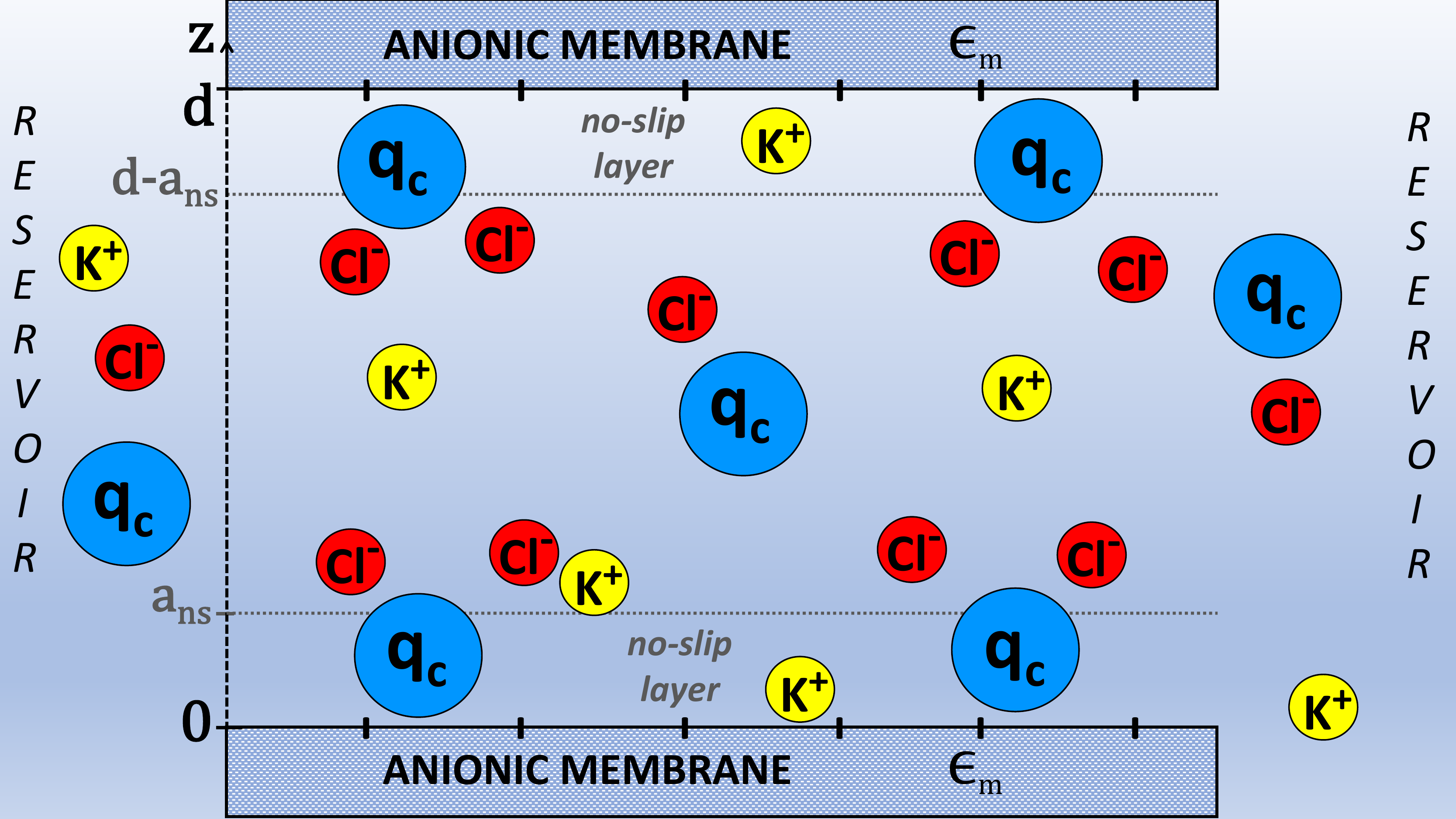}
\caption{(Color online) Schematic depiction of the slit pore with thickness $d$, length $L$, and width $w$. The slit located in an anionic membrane of permittivity $\e_{\rm m}$ confines the electrolyte ${\rm KCl}+{\rm X}^{q\ce}$ of permittivity $\e_{\rm w}$  including the multivalent ions $X^{q\ce}$ of valency $q\ce$. The charge configuration in the figure assumes $q\ce>0$.}
\label{fig1}
\end{figure}

The charge composition of the system is depicted in Fig.~\ref{fig1}. The electrolyte is composed of the monovalent salt KCl including ions with valencies $q_\pm=\pm1$ and bulk concentrations $n_{\pm\rm b}$, and a multivalent ion species with bulk concentration $n_{c\rm b}$ and valency $q\ce$ of arbitrary sign. The charge configuration in Fig.~\ref{fig1} corresponds to the specific case of multivalent cations ($q\ce>0$). The electrolyte is confined to a slit pore of thickness $d$ located in a membrane of dielectric permittivity $\e_{\rm m}$. The pore is composed of two anionic planes of surface charge density $-\sigma_{\rm m}$, lateral length $L$, and width $w$. Neglecting the finite size effects associated with the lateral wall boundaries where the slit is connected to an ion reservoir, the surface charge distribution is given by
\be\label{sc}
\sigma(\br)=-\sigma_{\rm m}\left[\delta(z)+\delta(d-z)\right].
\ee

The ions interact via the repulsive HC potential $w(\br-\br')$ and the electrostatic Coulomb potential $v\ce(\br,\br')$. The HC  potential is defined as $w(\br-\br')=\infty$ if $||\br-\br'||\leq2a_{\rm hc}$, and $w(\br-\br')=0$ for $||\br-\br'||>2a_{\rm hc}$, with the same HC radius $a_{\rm hc}$ assumed for all ionic species. The Coulomb potential is defined in terms of its inverse,
\be\label{cop}
v^{-1}\ce(\br,\br')=-\frac{k_{\rm B}T}{e^2}\nabla\e(\br)\nabla\delta(\br-\br'),
\ee
where $k_{\rm B}$ is the Boltzmann constant, $e$ the electron charge, the liquid temperature is $T=300$ K, and the dielectric permittivity profile reads 
\be\label{diel}
\e(\br)=\e_{\rm m}\left[\theta(-z)+\theta(z-d)\right]+\e_{\rm w}\theta(z)\theta(d-z),
\ee
with the dielectric permittivity of the solvent $\e_{\rm w}=80$.

By introducing two Hubbard-Stratonovich transformations associated with the fluctuating potentials $\phi(\br)$ and $\psi(\br)$ for the pairwise Coulombic and HC ion interactions, respectively, the grand-canonical partition function of the system can be recast as the functional integral~\cite{NetzHC}
\be\label{zg2}
Z_{\rm G}=\int\mathcal{D}\phi\mathcal{D}\psi\;e^{-\beta H[\phi,\psi]}.
\ee
In Eq.~(\ref{zg2}), the Hamiltonian functional reads $H[\phi,\psi]=H\s[\phi,\psi]+H\ce[\phi,\psi]$, where the solvent-implicit monovalent salt and multivalent ion components are defined as
\bea\label{ham2}
\beta H\s[\phi,\psi]&=&\frac{k_{\rm B}T}{2e^2}\int\mathrm{d}\br\e(\br)\left[\nabla\phi(\br)\right]^2-i\int\mathrm{d}\br\sigma(\br)\phi(\br)\nonumber\\
&&+\frac{1}{2}\int\mathrm{d}\br\mathrm{d}\br'\psi(\br)w^{-1}(\br-\br')\psi(\br')\nonumber\\
&&-\sum_{i=\pm}\int\mathrm{d}\br\;\hn_i(\br),\\
\label{ham3}
\beta H\ce[\phi,\psi]&=&-\int\mathrm{d}\br\;\hn\ce(\br).
\eea

The first three terms on the r.h.s. of Eq.~(\ref{ham2}) correspond respectively to the solvent free energy, the contribution from the fixed membrane charge $\sigma(\br)$, and the gaussian field contribution from the pairwise HC interactions. Then, the ion density term $\hn_i(\br)$ in Eqs.~(\ref{ham2})-(\ref{ham3}) is the contribution from the mobile salt charges ($i=\pm$) and multivalent ions ($i={\rm c}$). Therein, the fluctuating ion density of the species $i$ with fugacity $\Lambda_i$ reads 
\be
\hn_i(\br)=\Lambda_ie^{-V_i(\br)+iq_i\phi(\br)+i\psi(\br)},
\ee
where $V_i(\br)$ stands for the steric potential restricting the phase space accessible to the ions, and enabling the derivation of their average density from the grand potential $\beta\Omega_{\rm G}=-\ln Z_{\rm G}$ via the thermodynamic identity
\be\label{den1}
n_i(\br)=\frac{\delta\left[\beta\Omega_{\rm G}\right]}{\delta V_i(\br)}=\lan \hn_i(\br)\ran.
\ee
In Eq.~(\ref{den1}), the bracket notation designates the statistical average of the functional $F[\phi,\psi]$ over the fluctuations of the potentials $\phi(\br)$ and $\psi(\br)$, i.e.
\be
\label{av1}
\lan F[\phi,\psi]\ran=\frac{1}{Z_{\rm G}}\int \mathcal{D}\phi\mathcal{D}\psi\;e^{-\beta H[\phi,\psi]}F[\phi,\psi].
\ee
The steric constraint $a_{\rm st}<z<d-a_{\rm st}$ imposing to the ions the closest approach distance $a_{\rm st}$ to the slit walls will be taken as $e^{-V_i(\br)}=\theta(z-a_{\rm st})\theta(d-a_{\rm st}-z)$.

\subsection{Derivation of the electrostatic SCPB equation}


Based on the SD identity, a simplified version of the SCSD approach~\cite{Buyuk2020} will be derived. For the derivation of the SD Eq., one expresses first the variation of the partition function~(\ref{zg2}) by an infinitesimally small potential shift $\delta\phi(\br)$ as $\delta Z_{\rm G}=\int\mathcal{D}\phi\mathcal{D}\psi\;e^{-\beta H[\phi+\delta\phi,\psi]}-Z_{\rm G}$. In the remainder of the article, the notation will be simplified by omitting the potential dependence of the functionals. Expanding now the r.h.s. of the functional integral above at the linear order in $\delta\phi(\br)$, one obtains $\delta Z_{\rm G}=- \beta Z_{\rm G}\int\mathrm{d}\br\delta\phi(\br)\lan\delta H/\delta \phi(\br)\ran$. Then, one notes that the shift $\delta\phi(\br)$ in the same functional integral can be absorbed into the redefinition of the functional measure $\mathcal{D}\phi$. This implies the invariance of the partition function $Z_{\rm G}$ under the potential shift, i.e. $\delta Z_{\rm G}=0$. This finally yields the formally exact SD identity corresponding to the equation of state for the average potential, 
\be\label{av4}
\lan\frac{\delta H}{\delta \phi(\br)}\ran=0.
\ee


At this point, we introduce the SC treatment of the dilute multivalent ions equivalent to a low fugacity expansion of the grand potential~\cite{NetzSC}. First, by Taylor-expanding the SD Eq.~(\ref{av4}) together with Eq.~(\ref{av1}) at the linear order in the counterion density $\hn\ce(\br)$, one obtains 
\bea
\label{SD3}
\lan \frac{\delta H}{\delta \phi(\br)}\ran&=&\lan \frac{\delta H}{\delta \phi(\br)}\rans\\
&&-\left\{\lan \frac{\delta H\s}{\delta \phi(\br)}H\ce\rans-\lan \frac{\delta H\s}{\delta \phi(\br)}\rans\lan H\ce\rans\right\}=0.\nonumber
\eea
In Eq.~(\ref{SD3}), the average of the general functional $F[\phi,\psi]$ with the salt Hamiltonian~(\ref{ham2}) has been defined as
\be
\label{brs}
\lan F\rans=\frac{1}{Z\s}\int \mathcal{D}\phi\mathcal{D}\psi\;e^{-\beta H\s}F,
\ee
where the salt partition function reads $Z\s=\int \mathcal{D}\phi\mathcal{D}\psi\;e^{-\beta H\s}$. Moreover, at the same order $O\left(\hn\ce\right)$, the SC expansion of the ion density functions~(\ref{den1}) gives
\bea
\label{de1}
n_\pm(\br)&=&\lan \hn_\pm(\br)\rans+\int\mathrm{d}\brc\left[\lan \hn_\pm(\br)\hn\ce(\brc)\rans\right.\\
&&\left.\hspace{2.8cm}-\lan \hn_\pm(\br)\rans\lan\hn\ce(\brc)\rans\right],\nonumber\\
\label{de2}
n\ce(\br)&=&\lan \hn\ce(\br)\rans.
\eea
Inserting now the Hamiltonian components~(\ref{ham2})-(\ref{ham3}) into Eq.~(\ref{SD3}), the virial-expanded SD equation follows as
\bea\label{e1}
&&\frac{k_{\rm B} T}{e^2}\lan\nabla\e(\br)\nabla\phi(\br)\rans+i\sigma(\br)+i\sum_{i=\pm,{\rm c}}q_in_i(\br)\nonumber\\
&&=-\frac{k_{\rm B} T}{e^2}\int\mathrm{d}\brc\left[\lan\hn_{\rm c}(\brc)\nabla\e(\br)\nabla\phi(\br)\rans\right.\\
&&\hspace{1.8cm}\left.-\lan\hn_{\rm c}(\brc)\rans\lan\nabla\e(\br)\nabla\phi(\br)\rans\right].\nonumber
\eea


In order to evaluate the field averages in Eq.~(\ref{e1}) associated with the WC fluctuations of the monovalent salt around the MF PB solution, the salt Hamiltonian~(\ref{ham2}) will be approximated by the following Gaussian functional,
\bea\label{gaus}
H_{\rm s}&\approx&\frac{1}{2}\int_{\br,\br'}\left[\phi-i\phi\s\right]_\br G^{-1}(\br,\br')\left[\phi-i\phi\s\right]_{\br'}\nonumber\\
&&+\frac{1}{2}\int_{\br,\br'}\psi(\br)w^{-1}(\br-\br')\psi(\br').
\eea
Eq.~(\ref{gaus}) includes the inverse of the monovalent salt-dressed Debye-H\"{u}ckel (DH) Green's function,
\be\label{grop}
G^{-1}(\br,\br')=v^{-1}\ce(\br,\br')+\sum_{i=\pm}q_i^2n_{i\rm b} e^{-V_i(\br)}\delta(\br-\br'),
\ee
and the salt-screened average potential $\phi\s(\br)$ solving the SD Eq.~(\ref{e1}). Using now the identity
\be\label{in}
\int\mathrm{d}\br''G^{-1}(\br,\br'')G(\br'',\br')=\delta(\br-\br')
\ee
with Eqs.~(\ref{cop}) and~(\ref{grop}), the differential equation solved by the DH Green's function follows as
\be
\label{gr1}
\left[\nabla\e(\br)\nabla-\e(\br)\kappa^2(r)\right]G(\br,\br')=-\frac{e^2}{k_{\rm B}T}\delta(\br-\br').
\ee
In Eq.~(\ref{gr1}), we introduced the DH screening parameter
\be
\label{dh}
\kappa(r)=\kappa e^{-V_i(\br)}\;;\hspace{5mm}\kappa^2=4\pi\ell_{\rm B}\sum_{i=\pm}q_i^2n_{i\rm b}.
\ee
The solution of the kernel Eq.~(\ref{gr1}) in the planar geometry of the nanoslit is reported in Appendix~\ref{ap1}.

Computing the field averages in Eqs.~(\ref{de1})-(\ref{e1})  with the Hamitonian~(\ref{gaus}), the SCPB Eq.~(\ref{e1}) becomes
\bea
\label{e2}
&&\frac{k_{\rm B} T}{e^2}\nabla_\br\cdot\e(\br)\nabla_\br\left\{\phi\s(\br)+q\ce\int\mathrm{d}\brc n\ce(\brc)G(\brc,\br)\right\}\nonumber\\
&&+\sum_{i=\pm,c}q_in_i(\br)+\sigma(\br)=0,
\eea
with the ion densities 
\bea\label{deni}
n_\pm(\br)&=&\Lambda_\pm\;e^{-w(0)/2-V_\pm(\br)-q_\pm\phi\s(\br)-q_\pm^2G(\br,\br)/2}\nonumber\\
&&\hspace{1mm}\times\left\{1+\int\mathrm{d}\brc n\ce(\brc)f_\pm(\br,\brc)\right\},\\
\label{denc}
n\ce(\br)&=&\Lambda\ce\;e^{-w(0)/2-V\ce(\br)-q\ce\phi\s(\br)-q\ce^2G(\br,\br)/2},
\eea 
and the Mayer function
\be
\label{Mayer}
f_i(\br,\brc)=e^{- q_iq\ce G(\br,\brc)-w(\br-\brc)}-1.
\ee
Evaluating Eqs.~(\ref{deni})-(\ref{denc}) in the bulk reservoir where $\sigma(\br)=0$,  $\phi\s(\br)=0$, $n_i(\br)=n_{i{\rm b}}$ for $i=\{\pm,{\rm c}\}$, and $G(\br,\br')=G_{\rm b}(\br-\br')$, with the bulk Green's function $G_{\rm b}(r)=4\pi\ell_{\rm B}e^{-\kappa r}/r$, and plugging the resulting ion fugacities back into Eqs.~(\ref{deni})-(\ref{denc}),  the ion densities become at the order $O(n_{\rm cb})$
\bea
\label{d1}
n_\pm(\br)&=&n_{\pm\rm b}k_\pm(\br)\left[1+ n_{\rm cb}T_\pm(\br)\right],\\
\label{d2}
n\ce(\br)&=&n_{\rm cb}k\ce(\br).
\eea
In Eqs.~(\ref{d1})-(\ref{d2}), the function 
\be\label{br}
k_i(\br)=e^{-V_i(\br)-q_i\phi\s(\br)-q_i^2\delta G(\br)/2} 
\ee
corresponds to the bare ion density including the self-energy $\delta G(\br)=\left[G(\br,\br')-G\B(\br-\br')\right]_{\br'\to\br}$. Moreover, Eq.~(\ref{d1}) involves the many-body potentials incorporating the  direct interactions of the multivalent counterions with the salt ions,
\be
\label{T}
T_i(\br)=\int\mathrm{d}\br\ce\left[k\ce(\brc)f_i(\br,\brc)-f_{i\rm b}(\br-\brc)\right],
\ee
with the bulk limit of the Mayer function~(\ref{Mayer}),
\be
\label{Mayerb}
f_{i\rm b}(\br-\brc)=e^{- q_iq\ce G\B(\br-\brc)-w(\br-\brc)}-1.
\ee
One notes that the function~(\ref{T}) has the form of a non-uniform second virial coefficient $B_2$ weighted by the counterion density and renormalized by its bulk value~\cite{Hansen}.

Finally, combining Eqs.~(\ref{gr1}) and~(\ref{e2}), the SCPB Eq. takes the following integro-differential form
\be
\label{e3}
\frac{k_{\rm B} T}{e^2}\nabla_\br\cdot\e(\br)\nabla_\br\phi\s(\br)+\sum_{i=\pm}q_in_i(\br)+\sigma(\br)=-\frac{\kappa^2\s(\br)}{4\pi\ell_{\rm B}}\phi\ce(\br),
\ee
where we introduced the potential component induced by all multivalent counterions in the liquid,
\be
\label{pc}
\phi\ce(\br)=q\ce n_{\rm cb}\int\mathrm{d}^3\br\ce k\ce(\brc)G(\br,\brc).
\ee
One notes that in the bulk limit where $n_i(\br)=n_{i{\rm b}}$ and $\int_{\br\ce}G\B(\br-\br\ce)=q_+^2n_{+\rm b}+q_-^2n_{-\rm b}$, the bulk electroneutrality condition consistently follows from Eq.~(\ref{e3}) as
\be
\label{be}
q_+n_{+\rm b}+q_-n_{-\rm b}+q\ce n_{c\rm b}=0.
\ee

In order to determine the net average potential, in Eq.~(\ref{av1}), we set $F=\phi(\br)$ and linearize the result in the counterion density. One gets at the order $O\left(\hn\ce\right)$
\be
\label{mp1}
\lan \phi(\br)\ran=\lan \phi(\br)\rans-\left[\lan\phi(\br)H\ce\rans-\lan \phi(\br)\rans\lan H\ce\rans\right].
\ee
Evaluating the salt averages in Eq.~(\ref{mp1}), the real average potential $\Phi(\br)=-i\lan \phi(\br)\ran$ follows as
\be
\label{mp2}
\Phi(\br)=\phi\s(\br)+\phi\ce(\br).
\ee
Eq.~(\ref{mp2}) indicates that the net average potential is given by the superposition of the salt-dressed potential $\phi\s(\br)$ created by the membrane charge and satisfying the SCPB Eq.~(\ref{e3}), and the potential component~(\ref{pc}) induced by the multivalent counterions~\cite{rem1}.  In this work, the SCPB Eq.~(\ref{e3}) will be solved via  a perturbative expansion in terms of the counterion concentration $n_{\rm cb}$. The details of this solution scheme are reported in Appendix~\ref{per}.

Combining Eqs.~(\ref{e2}) and~(\ref{mp2}), the SCPB Eq.~(\ref{e3}) can be expressed in terms of the net potential $\Phi(\br)$ as
\be
\label{mp3}
\frac{k_{\rm B} T}{e^2}\nabla_\br\cdot\e(\br)\nabla_\br\Phi(\br)+\rho_{\rm ch}(\br)+\sigma(\br)=0.
\ee
In Eq.~(\ref{mp3}), the total mobile charge density reads
\be\label{st3}
\rho_{\rm ch}(\br)=\sum_{i=\pm,{\rm c}}q_in_i(\br),
\ee
with the bare partition function~(\ref{br}) taking the form $k_i(\br)=e^{-V_i(\br)-q_i\left[\Phi(\br)-\phi\ce(\br)\right]-q_i^2\delta G(\br)/2}$. The alternative form~(\ref{mp3}) of the SCPB Eq.~(\ref{e3}) is used below for the computation of the beyond-MF ionic current. 

\subsection{Computation of the ionic current}

The net current driven by an external voltage $\Delta V$ and pressure gradient $\Delta P$ through the nanoslit reads
\be
\label{st1}
I=we\int_{a_{\rm ns}}^{d-a_{\rm ns}}\mathrm{d}z\rho_{\rm ch}(z)u_i(z),
\ee
where the hydrodynamic no-slip radius $a_{\rm ns}$ corresponds to the characteristic thickness of the stagnant liquid layers composed of the multivalent ions strongly bound to the slit walls (see Fig.~\ref{fig1}). Moreover, the fluid velocity $u_i(z)=u_{{\rm T},i}+u\ce(z)$ is given by the superposition of the conductive velocity $u_{{\rm T},i}=\mu_i{\rm sign(q_i)}\Delta V/L$ characterized by the ionic mobility $\mu_i$~\cite{ionm}, and the convective velocity $u\ce(z)$ satisfying the Stokes Eq. $\eta u''\ce(z)+\Delta P/L+E\rho_{\rm ch}(z)=0$~\cite{Butt}, with the liquid viscosity $\eta=8.91\times10^{-4}$ Pa s, and the external electric field $E=\Delta V/L$. Solving the Stokes Eq. together with the relation $\rho_{\rm ch}(z)=-\Phi''(z)/4\pi\ell_{\rm B}$ following from the SCPB Eq.~(\ref{mp3}), and imposing the no-slip conditions $u\ce(a_{\rm ns})=0$ and $u\ce(d-a_{\rm ns})=0$, the convective velocity follows as
\bea
\label{st4}
u\ce(z)&=&\frac{\Delta P}{2\eta L}\left[(d-z)z-(d-a_{\rm ns})a_{\rm ns}\right]\\
&&+\mu_{\rm ep}E\left[\Phi(z)-\Phi(a_{\rm ns})\right],\nonumber
\eea
with the electrophoretic transport coefficient $\mu_{\rm ep}=e/(4\pi\ell_{\rm B}\eta)$. Finally, substituting Eq.~(\ref{st4}) into Eq.~(\ref{st1}), using the equality $\rho_{\rm ch}(z)=-\Phi''(z)/4\pi\ell_{\rm B}$, and performing integrations by parts, the net current becomes~\cite{Arellano2009}
\be
\label{st5}
I=K_{\rm p}\Delta P+K_{\rm v}\Delta V,
\ee
where the conductance components are
\bea
\label{st6}
K_{\rm p}&=&\frac{ew(d-2a_{\rm ns})}{4\pi\ell_{\rm B}L\eta}\left\{\int_{a_{\rm ns}}^{d-a_{\rm ns}}\frac{\Phi(z)\mathrm{d}z}{d-2a_{\rm ns}}-\Phi(a_{\rm ns})\right\},\\
\label{st7}
K_{\rm v}&=&\frac{we}{L}\int_{a_{\rm ns}}^{d-a_{\rm ns}}\mathrm{d}z\left\{\frac{e}{\eta}\left[\frac{\Phi'(z)}{4\pi\ell_{\rm B}}\right]^2
+\sum_{i=\pm,{\rm c}}\mu_i|q_i|n_i(z)\right\}.\nonumber\\
\eea
Eq.~(\ref{st6}) is the streaming conductance. Then, Eq.~(\ref{st7}) corresponds to the voltage-driven conductance. The first and second terms in the bracket of Eq.~(\ref{st7}) are the convective and conductive flow components, respectively.

\begin{figure}
\includegraphics[width=1.0\linewidth]{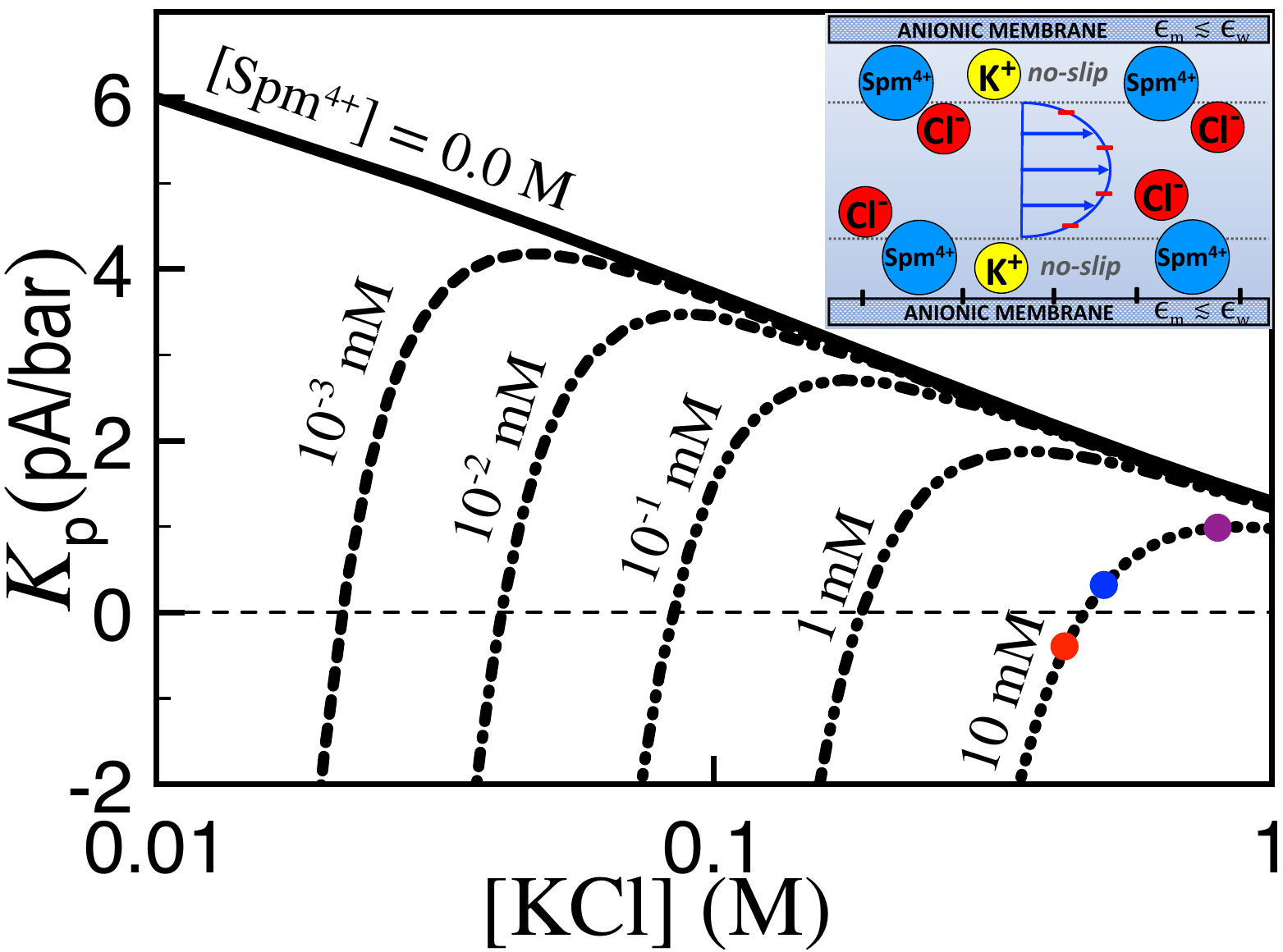}
\caption{(Color online) Streaming conductance~(\ref{st6}) versus the bulk KCl concentration at various ${\rm Spm}^{4+}$ concentration values ($q\ce=4$). The slit charge is $\sigma_{\rm m}=1.0$ ${\rm e}/{\rm nm}^2$~\cite{ExpHey}, the membrane permittivity $\e_{\rm m}=2$,  the pore radius $d=50$ nm, and the steric size $a_{\rm st}=0$. The no-slip and HC radii are $a_{\rm ns}=a_{\rm hc}=3$ {\AA}. The picture in the inset is a qualitative depiction of the current reversal mechanism.}
\label{fig2}
\end{figure}

\section{Results}

In this part, we apply the SCPB formalism to characterize the effect of strongly coupled multivalent charges on ion transport through nanochannels made in silicon (Si)-based insulator ($\e_{\rm m}\ll\e_{\rm w}$) and surface-coated dielectric membranes ($\e_{\rm m}\gg\e_{\rm w}$). We focus on the experimentally relevant case of slits with nanoscale thickness where the conductive component of Eq.~(\ref{st7}) largely dominates the convective part~\cite{Gillespie2013,the16}. The slit width and length are set to the values $w=50$ $\mu{\rm m}$ and $L=4.5$ mm of the transport experiments in Ref.~\cite{ExpHey}. 

\subsection{Streaming current reversal in charged slit pores}

We analyze here the experimentally observed multivalent ion-driven streaming current reversal~\cite{ExpHey}. Thus, the voltage is turned-off ($\Delta V=0$), and the membrane charge density and permittivity are set to the characteristic values $\sigma_{\rm m}=1.0$ ${\rm e}/{\rm nm}^2$ and $\e_{\rm m}=2$ of the silica nanochannels used in the transport experiments. Fig.~\ref{fig2} shows that in pure KCl solutions (solid black curve), salt decrement enhances the ${\rm K}^+$ adsorption and rises steadily the streaming conductance, i.e. $n_{+\rm b}\downarrow K_{\rm P}\uparrow$. However, if the KCl concentration is reduced in the presence of  ${\rm Spm}^{4+}$ molecules, $K_{\rm P}$ initially rises monotonically by following the pure salt curve, reaches a maximum where the effect of ${\rm Spm}^{4+}$ ions manifests itself, and starts dropping to switch from positive to negative at a characteristic KCl  concentration $n^*_{+\rm b}$. Moreover, the salt concentration at the current reversal rises with the amount of ${\rm Spm}^{4+}$, i.e. $n_{\rm cb}\uparrow n^*_{+\rm b}\uparrow$. These features agree qualitatively with the transport experiments~\cite{ExpHey} and earlier DFT studies~\cite{,Gillespie2011,Gillespie2013} conducted with trivalent cations.

%
\begin{figure}
	\includegraphics[width=1.0\linewidth]{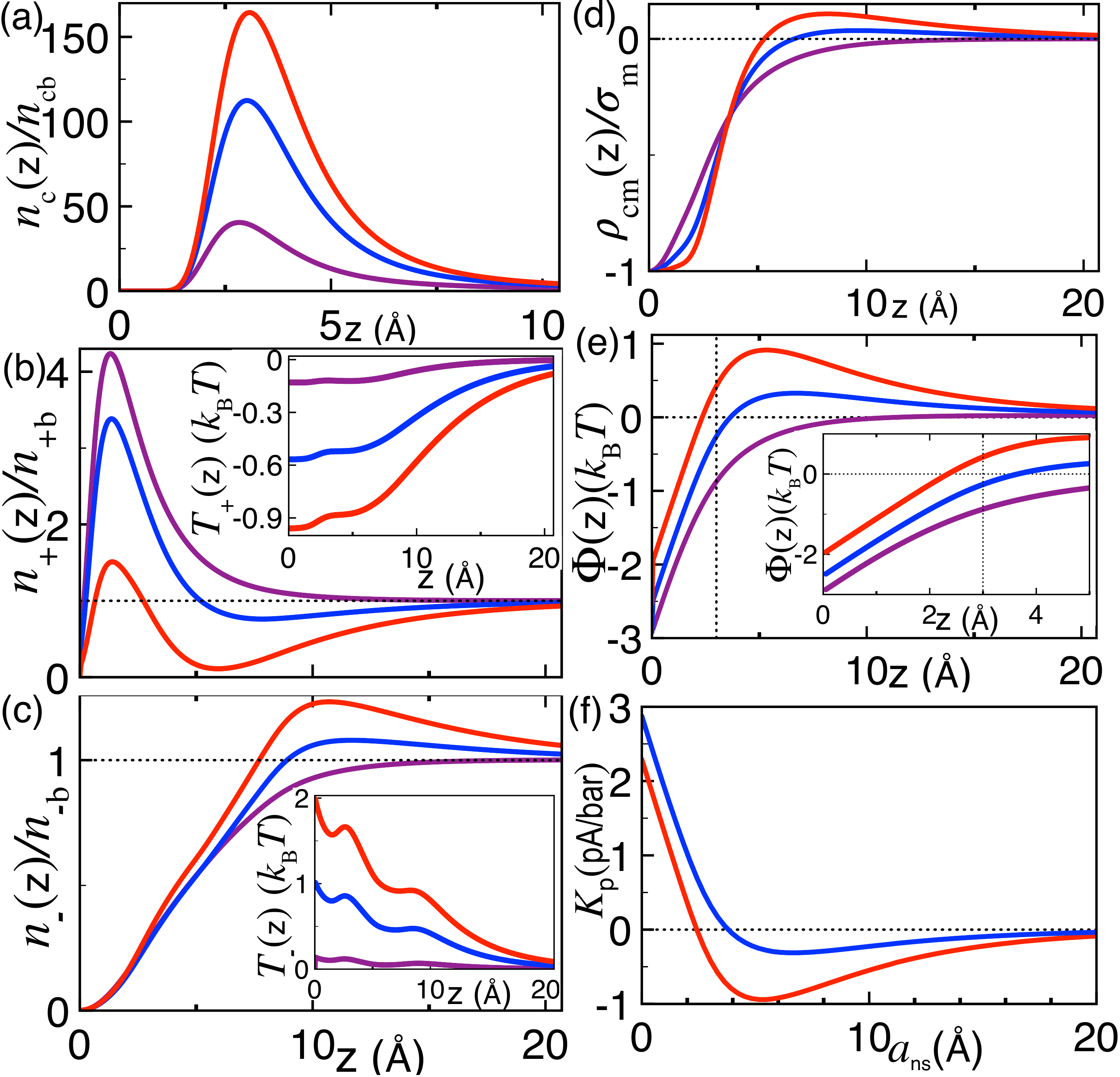}
	\caption{(Color online) (a) Multivalent counterion density~(\ref{d2}). (b)-(c) Monovalent ion densities~(\ref{d1}) (main plots) and the salt-multivalent interaction potentials~(\ref{T}) (insets). (d) Cumulative charge density~(\ref{st9}). (e) Net average potential~(\ref{mp2}). (f) Dependence of the conductance on the no-slip radius $a_{\rm ns}$. The salt concentration values of the curves corresponding to the dots of the same color in Fig.~\ref{fig2} are $n_{+\rm b}=0.8$ M (purple), $0.5$ M (blue), and $0.425$ M (red).}
	\label{fig3}
\end{figure}
The emergence of a negative current between the anionic slit walls is typically the signature of the membrane CI. In order to probe the actual correlation between these effects, we focus first on the charge configuration at the current reversal. Figs.~\ref{fig3}(a)-(c) display the ${\rm Spm}^{4+}$ and monovalent ion densities~(\ref{d1})-(\ref{d2}), and the virial function~(\ref{T}) at the salt concentrations of the dots with the same color in Fig.~\ref{fig2}. As the bulk KCl concentration is reduced from the positive current $K_{\rm P}>0$ (purple) to the negative current $K_{\rm P}<0$ regime (red), the suppression of the membrane charge screening drives further  ${\rm Spm}^{4+}$ ions into the slit, i.e. $n_{+\rm b}\downarrow n\ce(z)\uparrow$. This strengthens the ${\rm Spm}^{4+}-{\rm K}^+$ repulsion ($T_+(z)\downarrow$) and the ${\rm Spm}^{4+}-{\rm Cl}^-$ attraction ($T_-(z)\uparrow$), depleting the interfacial ${\rm K}^+$ ions ($n_+(z)\downarrow$) and amplifying the ${\rm Cl}^-$ density ($n_-(z)\uparrow$) above its bulk value. 
\begin{figure}
	\includegraphics[width=1.0\linewidth]{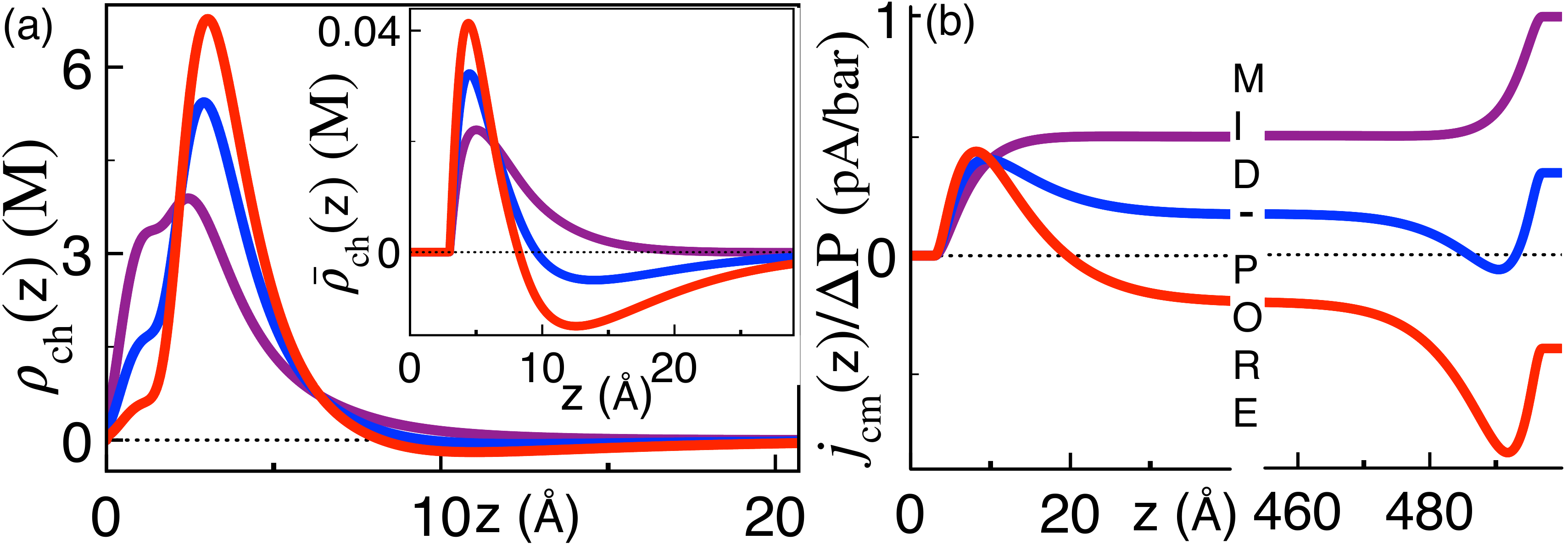}
	\caption{(Color online) (a) Charge density~(\ref{st3}) (main plot) and normalized charge flux $\bar{\rho}_{\rm ch}(z)=\rho_{\rm ch}(z)u\ce(z)/u\ce(d/2)$ (inset). (b) Cumulative current density~(\ref{st10}). Model parameters for each color are the same as in Fig.~\ref{fig3}.}
	\label{fig4}
\end{figure}

We investigate now the mechanism enabling these excess ${\rm Cl}^-$ ions to induce a negative current despite the abundance of the strongly cationic ${\rm Spm}^{4+}$ molecules. Fig.~\ref{fig3}(d) displays the cumulative charge density 
\be\label{st9}
\rho_{\rm cm}(z)\equiv\int_0^z\mathrm{d}z'\rho_{\rm ch}(z')-\sigma_{\rm m}=-\frac{\Phi'(z)}{4\pi\ell_{\rm B}},
\ee
where the second equality follows from the SCPB Eq.~(\ref{mp3}). One sees that the apparent like-charge ${\rm Cl}^-$ attraction and opposite-charge ${\rm K}^+$ exclusion by the anionic membrane manifests itself as the membrane CI. However, at the intermediate KCl concentration (blue) where CI is noticeable ($\rho_{\rm cm}(z)>0$), the conductance is not yet inverted ($K_{\rm P}>0$). This  indicates the absence of one-to-one mapping between CI and current reversal~\cite{rem2}. 

For an analytical insight into the actual causality between CI and current reversal, one should note that in nanofluidic experiments where one typically has $\kappa d\gg1$, the integral part of the conductance~(\ref{st6}) is negligible, i.e.
\be
\label{st8}
K_{\rm p}\approx-\frac{ew(d-2a_{\rm ns})}{4\pi\ell_{\rm B}L\eta}\Phi(a_{\rm ns}).
\ee
Eq.~(\ref{st8}) implies that current inversion requires the reversal of the \textit{zeta potential} $\Phi(a_{\rm ns})$ at the no-slip distance $z=a_{\rm ns}$ rather than the cumulative charge~(\ref{st9}) or field $\Phi'(z)$ at an arbitrary distance. Fig.~\ref{fig3}(e) indeed shows that the transition from positive (blue) to negative current (red) is accompanied with the shift of the potential reversal point into the no-slip region $z<3$ {\AA}. 

The requirement of potential inversion within the stagnant fluid layer indicates that current reversal is favored by the broadness of the no-slip region. Fig.~\ref{fig3}(f) confirms this point; as the no-slip distance $a_{\rm ns}$ increases from zero, the positive conductance drops and turns to negative before decaying due to the increasing overlap of the no-slip region with the ${\rm Cl}^-$ layer. In Fig.~\ref{fig4}, the electrohydrodynamic mechanism behind this peculiarity is illustrated in terms of the charge density~(\ref{st3}) and flux, and the integral of the flux corresponding to the cumulative current 
\be\label{st10}
j_{\rm cm}(z)=we\int_{a_{\rm ns}}^z\mathrm{d}z\rho_{\rm ch}(z')u\ce(z').
\ee
First, Fig.~\ref{fig4}(a) shows that the ${\rm Spm}^{4+}$ ions located in the stagnant fluid region $z\leq a_{\rm ns}$ and responsible for the strongly positive electrostatic charge density $\rho_{\rm ch}(z)$ do not contribute to the hydrodynamic charge flux $\bar\rho_{\rm ch}(z)$. Then, by comparing the maxima and minima in the main plot and the inset, one notes that the Poisseuille velocity~(\ref{st4}) rising steadily in the subsequent liquid region attenuates the positive charge contribution from the mobile ${\rm Spm}^{4+}$ ions at $z>a_{\rm ns}$, and enhances the anionic contribution of the next ${\rm Cl}^-$ layer where $\rho_{\rm ch}(z)<0$.

\begin{figure}
	\includegraphics[width=1.0\linewidth]{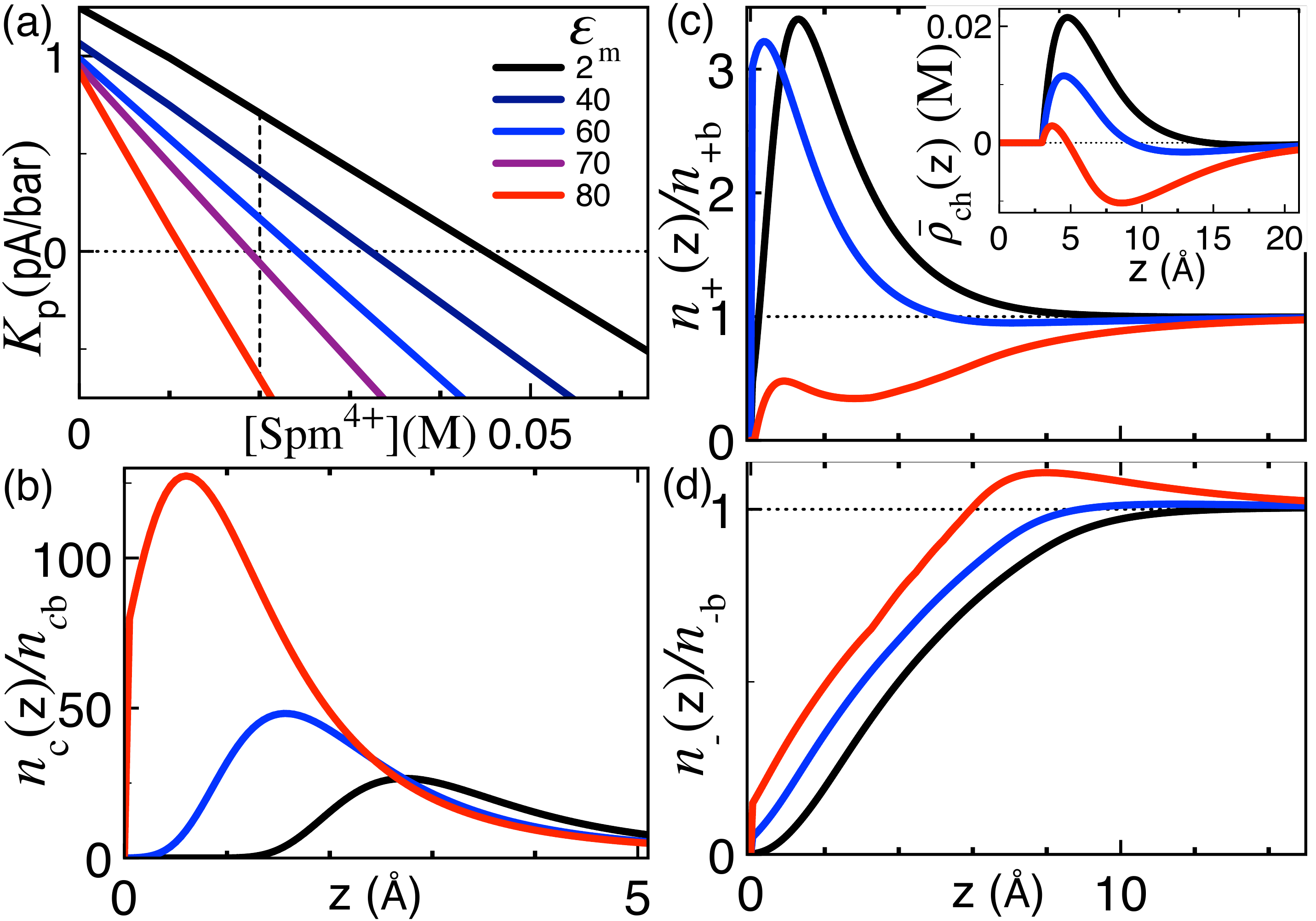}
	\caption{(Color online) (a) Streaming conductance~(\ref{st6}) versus the bulk ${\rm Spm}^{4+}$ concentration. (b) Local ${\rm Spm}^{4+}$ and (c)-(d) monovalent ion densities (main plots), and the normalized charge flux $\bar{\rho}_{\rm ch}(z)=\rho_{\rm ch}(z)u\ce(z)/u\ce(d/2)$ (inset) at the bulk ${\rm Spm}^{4+}$ concentration $n_{\rm cb}=0.02$ M (dashed line in (a)). The bulk salt density is $n_{\rm +b}=1.0$ M in all plots. The membrane permittivity $\e_{\rm m}$ for each color is reported in the legend of (a). The remaining model parameters are the same as in Fig.~\ref{fig2}.}
	\label{fig5}
\end{figure}

To summarize, the ${\rm Spm}^{4+}$ layer around each no-slip boundary acts as an effective positive surface charge attracting mobile ${\rm Cl}^-$ anions without fully contributing itself to the charge flow. In Fig.~\ref{fig3}(f), the emergence of the anionic current by the increase of the no-slip length originates precisely from the enlargement of this cation layer with reduced mobility. Fig.~\ref{fig4}(b) shows that as the salt decrement enhances the adsorption of these counterions by the slit, the mobile  ${\rm Cl}^-$ excess set by the counterions becomes large enough to invert the cumulative current $j_{\rm cm}(z)$ from positive (blue) to negative (red), generating a net anionic current  $I=j_{\rm cm}(z=d-a_{\rm ns})<0$ through the pore.  This effect is illustrated in the inset of Fig.~\ref{fig2}.


Thus, streaming current reversal is the collective effect of CI and the reduced interfacial liquid mobility. Fig.~\ref{fig5}(a) displays  the additional role of repulsive image charge forces in terms of the streaming conductance versus the ${\rm Spm}^{4+}$ concentration. The comparison of the curves at different membrane permittivities shows that the decrease of the permittivity at fixed ${\rm Spm}^{4+}$ concentration suppresses current reversal.  As a result, lower substrate permittivities lead to the occurrence of the current reversal at larger ${\rm Spm}^{4+}$ concentrations, i.e. $\e_{\rm m}\downarrow n^*_{\rm cb}\uparrow$.

\begin{figure}
	\includegraphics[width=1.0\linewidth]{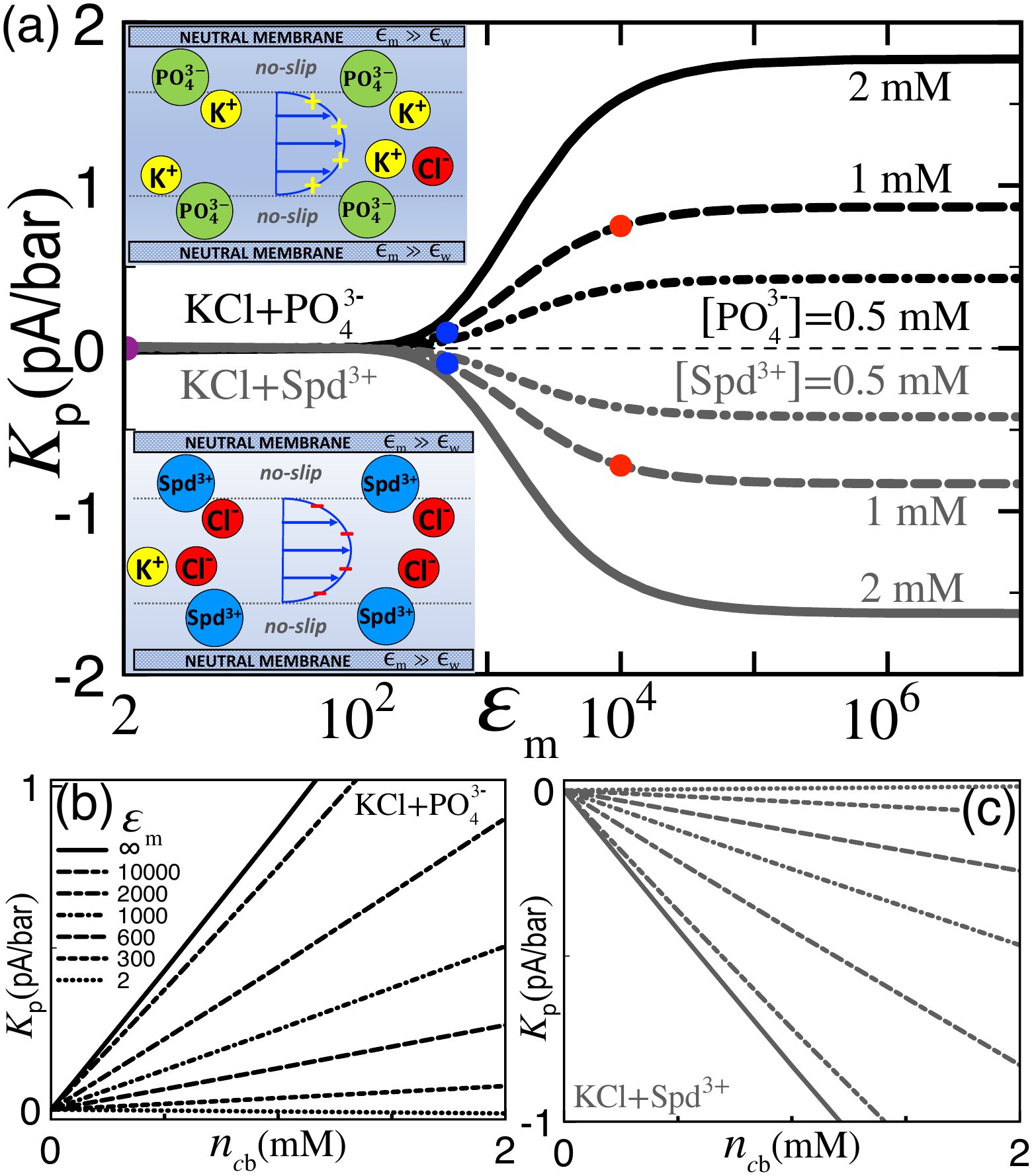}
	\caption{(Color online) (a) Polarization-induced streaming current against the permittivity of the neutral membrane and (b)-(c) the multivalent ion concentration in the mixed solutions ${\rm KCl}+{\rm PO}^{3-}_4$  (black curves) and ${\rm KCl}+{\rm Spd}^{3+}$ (grey curves). The insets in (a) depict the charge configuration at the origin of the current. The steric ion size and salt concentration are $a_{\rm st}=1.5$ {\AA} and $n_{\rm +b}=0.1$ M. The other parameters are the same as in Fig.~\ref{fig2}.}
	\label{fig6}
\end{figure}
Figs.~\ref{fig5}(b)-(d) show that the repulsive polarization forces amplified by the reduction of the substrate permittivity bring two major effects cancelling the current inversion. First, the image charge interactions reject the ${\rm Spm}^{4+}$ ions from the slit ($\e_{\rm m}\downarrow n\ce(z)\downarrow$), which reduces the ${\rm Cl}^-$ density and rises the ${\rm K}^+$ density ($n_-(z)\downarrow n_+(z)\uparrow$). Then, the repelled ${\rm Spm}^{4+}$ ions move from the no-slip layer towards the hydrodynamically mobile region. These two effects enhance the electrohydrodynamic weight of the ${\rm Spm}^{4+}$ and ${\rm K}^+$ cations with respect to the ${\rm Cl}^-$ anions, which turns the charge flux from negative back to positive (see the inset) and suppresses the current reversal. Next, we consider the opposite case of attractive polarization forces emerging in the giant permittivity regime of surface-coated membrane nanoslits.

\subsection{Polarization-induced streaming currents and ion separation through surface-coated pores}
\label{stpol}

The surface coating of low permittivity membranes by CNTs is known to enhance the dielectric permittivity of the substrate into the giant permittivity regime $\e_{\rm m}\gg\e_{\rm w}$ characterized by attractive polarization forces~\cite{Coating}. Here, we reveal a new mechanism of streaming current generation triggered by the action of these forces on the multivalent ions of the electrolyte mixture.

\begin{figure}
	\includegraphics[width=1.0\linewidth]{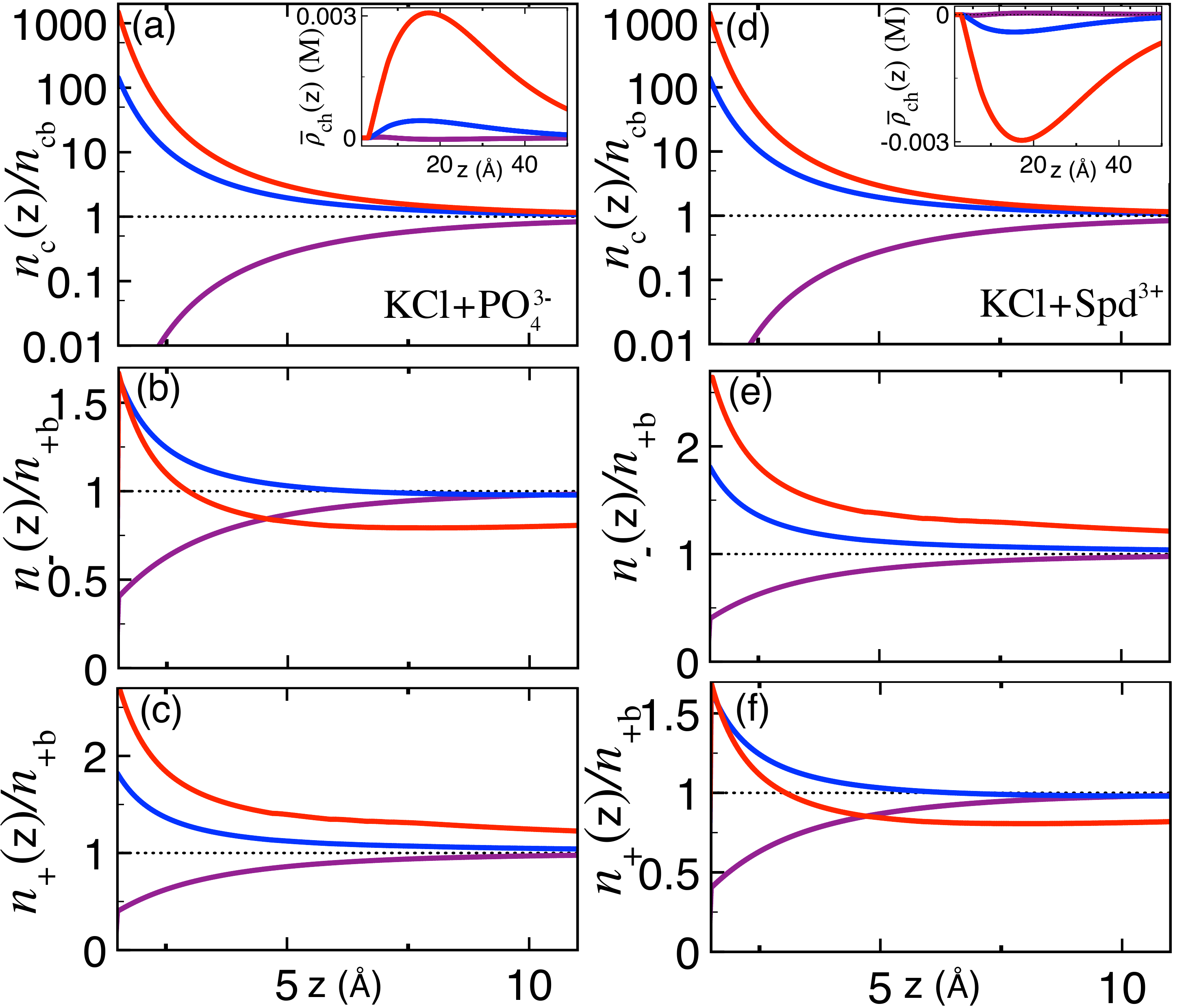}
	\caption{(Color online) (a) Multivalent counterion density~(\ref{d2}) (main plot) together with the normalized charge flux $\bar{\rho}_{\rm ch}(z)=\rho_{\rm ch}(z)u\ce(z)/u\ce(d/2)$ (inset), and (b)-(c) the monovalent ion densities~(\ref{d1}) for the ${\rm KCl}+{\rm PO}^{3-}_4$ mixture confined to a neutral slit pore. The permittivity values corresponding to the dots of the same color in Fig.~\ref{fig6} are $\e_{\rm m}=2$ (purple), $500$ (blue), and $10^4$ (red). (d)-(f) display the equivalent plots for the ${\rm KCl}+{\rm Spd}^{3+}$ liquid.}
	\label{fig7}
\end{figure}
Fig.~\ref{fig6}(a) displays the permittivity dependence of the streaming conductance in neutral slits ($\sigma_{\rm m}=0$) containing the electrolyte mixtures ${\rm KCl}+{\rm PO}^{3-}_4$  (top plot) and ${\rm KCl}+{\rm Spd}^{3+}$ (bottom plot) with the trivalent phosphate (${\rm PO}^{3-}_4$) anions and spermidine (${\rm Spd}^{3+}$) cations. In a neutral slit confining a pure ${\rm KCl}$ solution, due to the symmetric charge partition, the local charge density~(\ref{st3}) and net current~(\ref{st1}) would vanish.  However, Fig.~\ref{fig6}(a) shows that in the presence of trivalent ions, the rise of the membrane permittivity from the insulator $\e_{\rm m}\ll\e_{\rm w}$ to the giant permittivity regime $\e_{\rm m}\gg\e_{\rm w}$ amplifies the streaming conductance from a vanishingly small magnitude to a substantially positive value for ${\rm PO}^{3-}_4$ anions, and a negative value for ${\rm Spd}^{3+}$ cations. Moreover, in the metallic regime $\e_{\rm m}\gtrsim10^5$, the current amplitude saturates at a limiting value roughly proportional to the multivalent ion concentration. This point is explicitly illustrated in Figs.~\ref{fig6}(b)-(c): for $\e_{\rm m}>\e_{\rm w}$, added ${\rm PO}^{3-}_4$ anions (${\rm Spd}^{3+}$ cations) into the ${\rm KCl}$ solution activate a positive (negative) conductance whose magnitude rises with the multivalent ion density ($K_{\rm p}\propto n_{\rm cb}$) and the membrane permittivity ($\e_{\rm m}\uparrow |K_{\rm p}|\uparrow$).

Thus, in the giant permittivity regime of the surface-coated pores, the attractive polarization forces coupled to the multivalent ions can solely set a counterion current whose sign and strength can be tuned by the type and amount of the added multivalent charges. This effect is the key prediction of our work. Fig.~\ref{fig7} illustrates the mechanism driving this current in terms of the mono- and multivalent ion densities, and the normalized charge flux for the ${\rm KCl}+{\rm PO}^{3-}_4$ solution (left) and the ${\rm KCl}+{\rm Spd}^{3+}$ mixture (right). The permittivity value for each curve corresponds to the dots of the same color in Fig.~\ref{fig6}(a). 
\begin{figure}
	\includegraphics[width=1.0\linewidth]{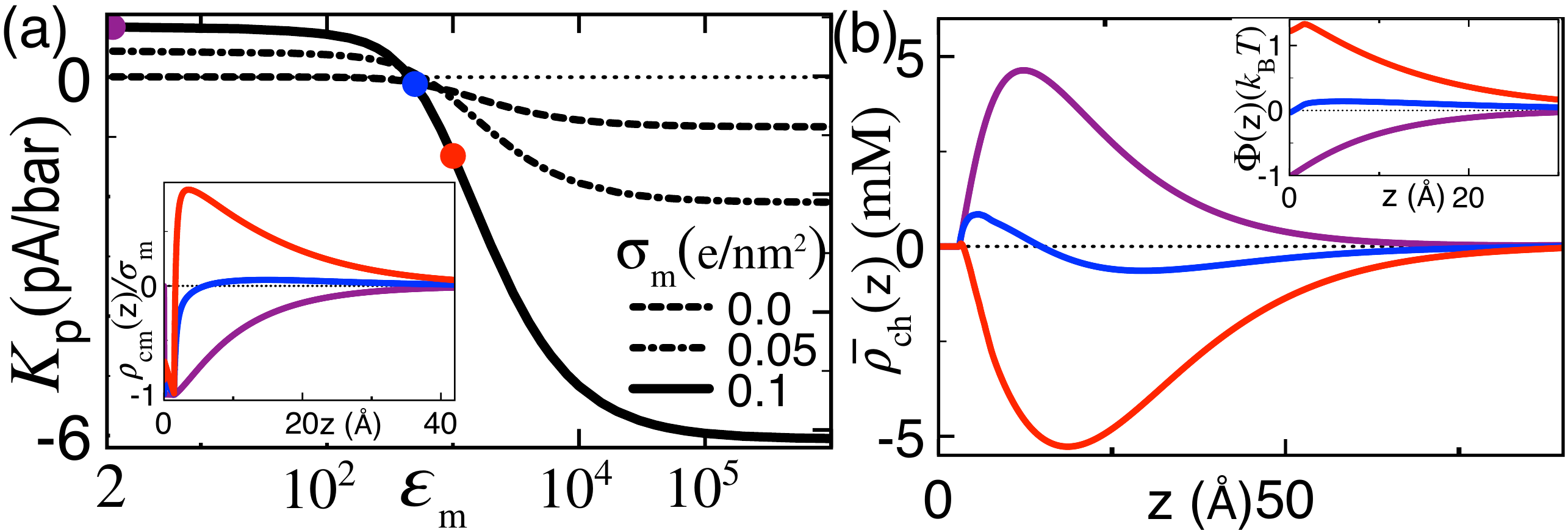}
	\caption{(Color online) (a) Permittivity dependence of the streaming conductance at different anionic surface charges $\sigma_{\rm m}$ (main plot) and the cumulative charge density~(\ref{st9}) (inset) for the ${\rm KCl}+{\rm Spd}^{3+}$ liquid. (b) Normalized charge flux (main plot) and the average potential~(\ref{mp2}) (inset). The membrane permittivities of the curves corresponding to the dots of the same color in (a) are $\e_{\rm m}=2$ (purple), $500$ (blue), and $10^3$ (red). The ion concentrations are $n_{\rm +b}=0.1$ M and $n_{\rm cb}=1.0$ mM. The other parameters are the same as in Fig.~\ref{fig2}.}
	\label{fig8}
\end{figure}

The plots indicate that the increment of the membrane permittivity from the repulsive (purple) to the attractive image charge regime $\e_{\rm m}=500$ (blue) switches the configuration of all ionic species from the exclusion ($n_{\pm,{\rm c}}(z)<n_{\pm,{\rm cb}}$) to the excess state ($n_{\pm,{\rm c}}(z)>n_{\pm,{\rm cb}}$). However, according to Fig.~\ref{fig6}(a), the substantial amplification of the pore conductance by polarization forces occurs in the subsequent permittivity regime $\e_{\rm m}\gtrsim500$. Indeed, Fig.~\ref{fig7} shows that due to the quadratic dependence of the attractive image charge interactions on the ion valency (see Eqs.~(\ref{deni})-(\ref{denc})), the multivalent ions experience a significantly stronger interfacial adsorption than the monovalent salt ions. In the permittivity regime $\e_{\rm m}\gtrsim500$ where this asymmetry becomes pronounced,  the salt-multivalent ion interactions take over the salt-image charge interactions. Consequently, the ${\rm PO}^{3-}$ (${\rm Spd}^{3+}$) adsorption attracts further ${\rm K}^+$ (${\rm Cl}^-$) ions but excludes the ${\rm Cl}^-$ (${\rm K}^+$) ions from the slit (red curves). The insets in Figs.~\ref{fig7}(a) and (d) show that this results in a charge flux opposite to the sign of the added multivalent charges, explaining the counterion current displayed in Fig.~\ref{fig6}. The corresponding charge separation mechanism is schematically depicted in the insets of Fig.~\ref{fig6}.
\begin{figure*}
\includegraphics[width=1.0\linewidth]{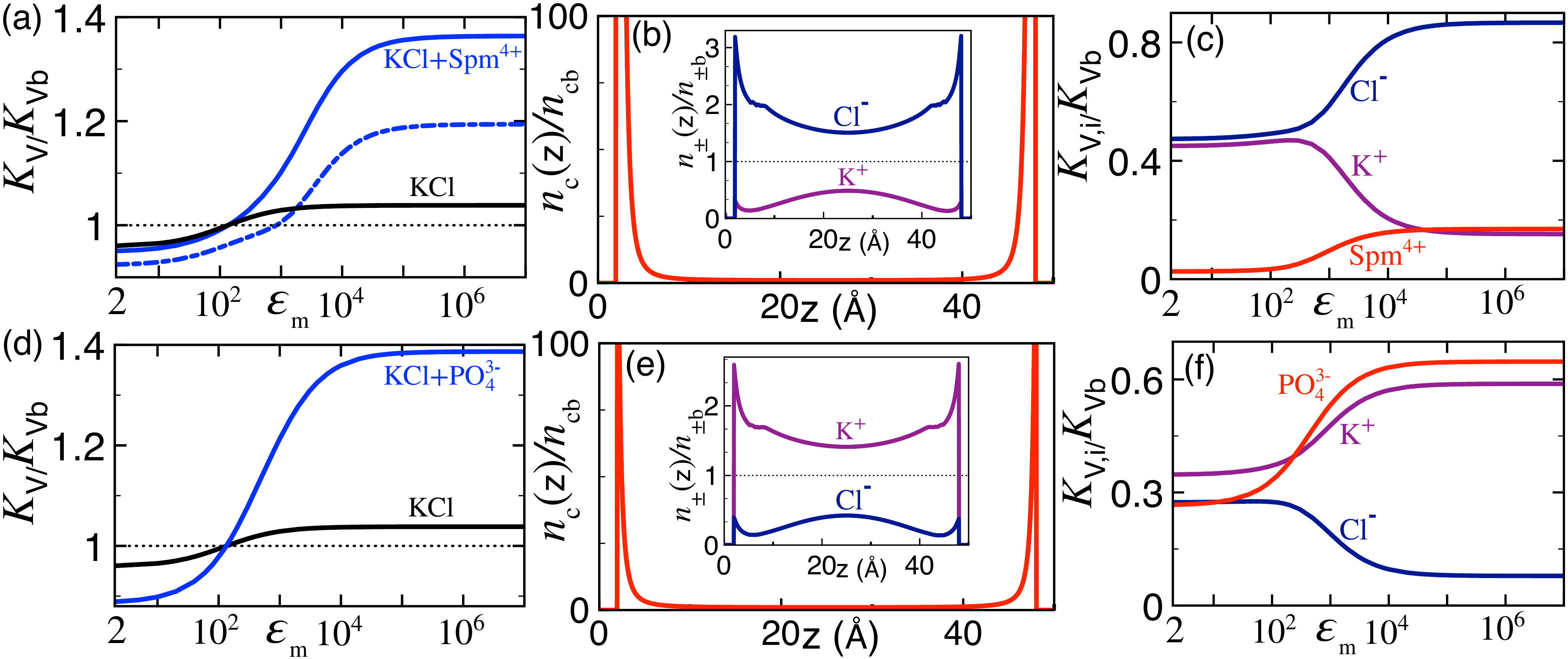}
\caption{(Color online) (a) Dielectric permittivity dependence of the voltage-driven conductance~(\ref{st7}) for the ${\rm KCl}$ solution (black curve) and the electrolyte mixture ${\rm KCl}+{\rm Spm}^{4+}$ (solid blue curve). (b) ${\rm Spm}^{4+}$ (main plot) and monovalent salt densities (inset) at the membrane permittivity $\e_{\rm m}=10^7$. (c) Conductance components~(\ref{concomp}) of separate species. The dashed blue curve in (a) displays the ${\rm KCl}+{\rm Spm}^{4+}$ conductance without the ${\rm Spm}^{4+}$ component, i.e. $(K_{\rm V}-K_{\rm V,{\rm c}})/K_{\rm Vb}$. The neutral pore size is $d=5$ nm, the steric ion size $a_{\rm st}=2.0$ {\AA}, and the ion concentrations are $n_{\rm +b}=0.1$ M and $n_{\rm cb}=0.5$ mM. (d)-(f) display the equivalent plots for the ${\rm KCl}+{\rm PO}^{3-}_4$ mixture with $n_{\rm cb}=10.0$ mM. The remaining parameters are the same as in Fig.~\ref{fig2}.}
\label{fig9}
\end{figure*}

Considering the difficulty in modifying externally the sign and density of the nanopore surface charges,  the polarization-induced streaming current presents itself as a highly relevant prediction for nanofluidic ion separation purposes. Figs.~\ref{fig8}(a)-(b) illustrate the extension of this mechanism to membranes with a finite anionic surface charge. One sees that in negatively charged slits, the increase of the membrane permittivity from the insulator to the giant permittivity regime leads to the pore CI (see the insets), and the reversal of the charge flux and conductance from a finite positive to a negative value (see the main plots). It is noteworthy that the activation of surface CI by the exclusive effect of the attractive polarization forces has been previously demonstrated in the MC simulations of multivalent electrolytes in contact with planar conductors~\cite{Wang2016}. This partial overlap of the results is a relevant numerical support of the polarization-induced current generation mechanism revealed herein.

\subsection{Ion separation by attractive polarization forces in voltage-driven charge transport}

The voltage-induced charge transport driven by the individual ion conductivities radically differs from the pressure-driven convective transport process studied in Sec.~\ref{stpol}. We probe here the effect of the attractive coupling between the polarization forces and the multivalent ions on the ionic composition of voltage-driven currents. We consider a neutral slit of small thickness $d=5$ nm where  deviations from the bulk transport behavior can be easily observed. In order to focus on the ion separation regime, the bulk multivalent ion concentration will be set to a sufficiently large value for the monovalent coions to be quasi-totally depleted from the slit.


Fig.~\ref{fig9}(a) displays the dielectric permittivity dependence of the conductance~(\ref{st7}) rescaled by its bulk value 
\be\label{conb}
K_{\rm Vb}=\frac{we}{L}(d-2a_{\rm ns})\sum_{i=\pm,{\rm c}}\mu_i|q_i|n_{i{\rm b}}
\ee
for the KCl and ${\rm KCl}+{\rm Spm}^{4+}$ solutions. As the permittivity rises from  the insulator to the metallic limit, the multivalent ion effects come into play in the giant permittivity regime $\e_{\rm m}\gtrsim10^3$ where the ${\rm KCl}+{\rm Spm}^{4+}$ conductance (solid blue curve) experiences a significantly stronger enhancement than the KCl conductance (black). In Fig.~\ref{fig9}(a), we reported as well the reduced conductance $(K_{\rm V}-K_{\rm V,{\rm c}})/K_{\rm Vb}$ excluding the ${\rm Spm}^{4+}$ contribution (dashed curve), with the conductance components of the separate ionic species $i=\{\pm,{\rm c}\}$ defined as
\be\label{concomp}
K_{{\rm V},i}=\frac{we}{L}\mu_i|q_i|\int_{a_{\rm ns}}^{d-a_{\rm ns}}\mathrm{d}zn_i(z).
\ee
One sees that upon the increment of the membrane permittivity from $\e_{\rm m}\sim2$ to $\e_{\rm m}\gg\e_{\rm w}$,  the reduced and total conductances are amplified by comparable amounts. This suggests that the principle role played by the ${\rm Spm}^{4+}$ cations on the current enhancement is electrostatic. 

The primarily electrostatic effect of the ${\rm Spm}^{4+}$ ions on the current amplification is corroborated in Figs.~\ref{fig9}(b)-(c). According to Fig.~\ref{fig9}(b), for $\e_{\rm m}\gg\e_{\rm w}$, the ${\rm Spm}^{4+}$ cations strongly adsorbed by the pore walls exclude the ${\rm K}^+$ cations and attract the ${\rm Cl}^-$ anions. Fig.~\ref{fig9}(c) shows that beyond the permittivity value $\e_{\rm m}\sim10^3$, this leads to the suppression of the ${\rm K}^+$ conductivity ($\e_{\rm m}\uparrow K_{{\rm V},+}\downarrow$), and the enhancement of the ${\rm Cl}^-$ conductivity ($\e_{\rm m}\uparrow K_{{\rm V},-}\uparrow$) responsible for the amplification of the ${\rm KCl}+{\rm Spm}^{4+}$ conductance observed in Fig.~\ref{fig9}(a). Although attractive polarization forces equally amplify the ${\rm Spm}^{4+}$ conductance (red curve in Fig.~\ref{fig9}(c)), due to the low bulk ${\rm Spm}^{4+}$ density, the former is largely dominated by the enhanced ${\rm Cl}^-$ component. Thus, in the giant permittivity regime of surface-coated pores, added ${\rm Spm}^{4+}$ ions induce an electrostatic charge discrimination and generate a ${\rm Cl}^-$-rich current without making themselves a substantial conductive contribution to charge transport.

Fig.~\ref{fig9}(d) shows that the addition of the trivalent ${\rm PO}^{3-}_4$ anions to the KCl solution leads to a quantitatively comparable enhancement of the total conductance, which is partly driven by the direct ${\rm PO}^{3-}_4$-${\rm K}^+$  interactions. Namely, in the regime $\e_{\rm m}\gg\e_{\rm w}$,  the adsorbed ${\rm PO}^{3-}_4$ ions setting an anionic surface charge exclude the ${\rm Cl}^-$ anions and attract the ${\rm K}^+$ cations (see Fig.~\ref{fig9}(e)). At $\e_{\rm m}\gtrsim10^3$, this reduces the ${\rm Cl}^-$ conductivity and enhances the ${\rm K}^+$ conductivity, i.e. $\e_{\rm m}\uparrow K_{{\rm V},-}\downarrow K_{{\rm V},+}\uparrow$ (see  Fig.~\ref{fig9}(f)). However, in contrast with the previous case of ${\rm Spm}^{4+}$ charges, the ${\rm PO}^{3-}_4$ ions of lower valency and larger bulk concentration also bring a sizable conductive contribution to the increment of the total current; one notes that in the giant permittivity regime, the ${\rm PO}^{3-}_4$ conductance becomes even higher than the ${\rm K}^+$ conductance. Hence the separate electrostatic and conductive contributions from the ${\rm PO}^{3-}_4$ ions lead to a voltage-driven current equally rich in the ${\rm K}^+$  and ${\rm PO}^{3-}_4$  species.

\section{Conclusions} 

The comprehension of charged systems characterized by the omnipresence of composite electrolytes requires the formulation of electrostatic theories able to take into account mixed interaction strengths. Via the self-consistent incorporation of dilute multivalent ions into the PB equation, we developped a theoretical formalism accounting for the electrostatic and HC interactions of WC monovalent salt and SC multivalent ions. By combining the corresponding SCPB formalism with the Stokes equation, we investigated the impact of SC electrostatics on nanofludic charge transport through nanoslits fabricated in low permittivity silica and surface-coated dielectric membranes.

The SCPB theory was applied to characterize the emergence of negative streaming currents in strongly anionic silica nanochannels~\cite{Heyden2005}. This exotic transport behavior is driven by the collective effect of the multivalent counterions activating the membrane CI and bringing ${\rm Cl}^-$ ions into the slit, and the no-slip zone reducing the contribution from these interfacial counterions to the net current. As a result of this cooperative mechanism, the like-charge current formation requires the inversion of the average potential {\it within the no-slip region}.

We have also probed SC effects on the pressure-driven transport properties of surface-coated nanopores located in the giant permittivity regime $\e_{\rm m}\gg\e_{\rm w}$ associated with attractive polarization forces~\cite{Coating}. Under the effect of these forces, added multivalent ions result in a charge separation and generate a {\it counterion current} through interfacially neutral slits. This key prediction can be readily verified by transport experiments and the underlying electrohydrodynamic mechanism can be beneficial to ion separation and water purification technologies. 

In addition, we found that if the nanoslit carries anionic surface charges, the increment of the membrane permittivity from the insulator to the giant permittivity regime can solely trigger the pore CI and reverse the streaming current from positive to negative. Thus, the polarization-driven current generation mechanism is supported by MC simulations where the activation of CI by the exclusive effect of the attractive image-charge interactions has been previously observed~\cite{Wang2016}. 

In the case of voltage-driven transport through neutral dielectric slits, charge separation by quadrivalent ${\rm Spm}^{4+}$ molecules generates a single-species current rich in ${\rm Cl}^-$. However, if the charge separation is achieved by the addition of trivalent ${\rm PO}^{3-}_4$ ions, their higher concentration resulting from their lower valency leads to a voltage-driven current equally rich in the ${\rm PO}^{3-}_4$ and ${\rm K}^+$ species. Thus, in the charge separation phase, the weight of the multivalent ions in the current drops inversely with their valency.

The consequences of the polarization-induced charge separation on the field-driven polymer translocation~\cite{Buyuk2018} will be explored in an upcoming work. Moreover, in Appendix~\ref{dri}, the SCPB formalism was shown to provide a self-consistent route to the dressed-ion theory of Refs.~\cite{Podgornik2010,Podgornik2011}. In the light of this exact mapping, the comparative confrontation of these formalisms with intensive MC simulations will help to identify the validity regime of the underlying approximations.

\smallskip
\appendix

\section{Green's function in slit pores}
\label{ap1}

We report here the solution of the DH Eq.~(\ref{gr1}) in the slit pore geometry. Due to the plane symmetry, the Green's function can be Fourier expanded as
\bea\label{Gr}
G(\br,\br')&=&\int\frac{\mathrm{d}^2\bk}{4\pi^2}e^{i\bk\cdot\left(\br_\pa-\br'_\pa\right)}\tG(z,z';k).
\eea
Injecting the expansion~(\ref{Gr}) into Eq.~(\ref{gr1}), and solving the resulting kernel equation with the dielectric permittivity profile~(\ref{diel}) and the closest ionic approach distance $a_{\rm st}$ to the planar walls, in the region $a_{\rm st}\leq z,z'\leq d-a_{\rm st}$, the Fourier-transformed Green's function follows as
\bea\label{frGr}
\tilde G(z,z';k)&=&\tilde G_{\rm b}(z-z';k)\\
&&+\frac{2\pi\ell_{\rm B}\Delta}{p\left[1-\Delta^2e^{-2p(d-2a_{\rm st})}\right]}\nonumber\\
&&\hspace{3mm}\times\left[e^{-p(z+z'-2a_{\rm st})}+e^{-p(2d-z-z'-2a_{\rm st})}\right.\nonumber\\
&&\hspace{7mm}\left.+2\Delta e^{-2p(d-2a_{\rm st})}\cosh\left(p|z-z'|\right)\right],\nonumber
\eea
with the bulk component $\tilde G_{\rm b}(z-z';k)=2\pi\ell_{\rm B}e^{-p|z-z'|}/p$, and the parameters $p=\sqrt{\kappa^2+k^2}$, $\Delta=(p-\eta k)/(p+\eta k)$, 
\bea
\eta=\frac{1-\Delta_0e^{-2ka_{\rm st}}}{1+\Delta_0e^{-2ka_{\rm st}}},
\eea
and $\Delta_0=(\e_{\rm w}-\e_{\rm m})/(\e_{\rm w}+\e_{\rm m})$~\cite{netzvdw,Buyuk2011}. Consequently, the self-energy $\delta G(\br)=\left[G(\br,\br')-G\B(\br-\br')\right]_{\br'\to\br}$ becomes
\bea
\label{sep}
\delta G(z)&=&\ell_{\rm B}\int_0^\infty\frac{\mathrm{d}kk}{p}\frac{\Delta}{1-\Delta^2e^{-2p(d-2a_{\rm st})}}\\
&&\hspace{0mm}\times\left[e^{-2p(z-a_{\rm st})}+e^{-2p(d-a_{\rm st}-z)}+2\Delta e^{-2p(d-2a_{\rm st})}\right].\nonumber
\eea

\section{Perturbative solution of the SCPB Eq.}
\label{per}

This appendix explains the perturbative solution of the SCPB Eq.~(\ref{e3}) via its virial expansion in terms of the multivalent ion concentration.

\subsection{Virial expansion of the SCPB Eq.~(\ref{e3})}
\label{vir}

The perturbative expansion of the SCPB Eq.~(\ref{e3}) will be carried-out by splitting first the potential satisfying this equation into a purely salt-dressed component $\phi_0(\br)$ and a correction potential $\phi_1(\br)$ associated with the scattering of the salt ions by the multivalent charges,
\be\label{sp}
\phi\s(\br)=\phi_0(\br)+\phi_1(\br).
\ee
In order to simplify the notation, from now on, we set $q_\pm=\pm1$. At the next step, we insert the electroneutrality condition~(\ref{be}) and Eq.~(\ref{sp}) into Eqs.~(\ref{d1})-(\ref{d2}). Taylor-expanding the result in terms of the counterion concentration $n_{\rm cb}$ and the perturbative correction potential $\phi_1(\br)$, at the order $O\left(n_{\rm cb}\right)$, the ion densities become 
\bea
\label{pd1}
n_+(\br)&=&n_{+\rm b}k_{{\rm s}+}(\br)\left[1-\phi_1(\br)+n_{\rm cb}T_+(\br)\right],\\
\label{pd2}
n_-(\br)&=&n_{+\rm b}k_{{\rm s}-}(\br)\left[1+\phi_1(\br)+n_{\rm cb}T_-(\br)\right]\nonumber\\
&&+q\ce n_{\rm cb}k_{{\rm s}-}(\br),\\
\label{pd3}
n\ce(\br)&=&n_{\rm cb}k_{\rm sc}(\br),
\eea
where $k_{{\rm s}i}(\br)=e^{-V_i(\br)-q_i\phi_0(\br)-q_i^2\delta G(\br)/2}$ and
\bea
\label{Ts}
T_i(\br)&=&\int\mathrm{d}\br\ce\left[k_{{\rm sc}}(\brc)f_i(\br,\brc)-f_{i\rm b}(\br-\brc)\right]+O\left(n_{\rm cb}\right)\nonumber\\
\eea
for $i=\{\pm,{\rm c}\}$. The implementation of the functions~(\ref{Ts}) in the slit pore geometry is explained in Appendix~\ref{tslit}.

Substituting now Eqs.~(\ref{pd1})-(\ref{pd3}) into the SCPB Eq.~(\ref{e3}), and linearizing the result in terms of the concentration $n_{\rm cb}$, one gets two differential equations satisfied by each potential component in Eq.~(\ref{sp}). The first equation is the standard PB equation augmented by the ionic self-energy $\delta G(\br)$~\cite{Buyuk2010},
\be
\label{e4}
\frac{k_{\rm B} T}{e^2}\nabla_\br\cdot\e(\br)\nabla_\br\phi_0(\br)-2n_{+\rm b}k_0(\br)\sinh\left[\phi_0(\br)\right]+\sigma(\br)=0,
\ee
with the auxiliary function $k_0(\br)=e^{-V_i(\br)-\delta G(\br)/2}$. The second identity is a second order linear differential equation for the counterion correction potential $\phi_1(\br)$,
\bea
\label{e6}
&&\frac{k_{\rm B} T}{e^2}\nabla_\br\cdot\e(\br)\nabla_\br\phi_1(\br)-2n_{+\rm b}k_0(\br)\cosh\left[\phi_0(\br)\right]\phi_1(\br)\nonumber\\
&&=-n_{\rm cb}J(\br),
\eea
with the source function
\bea
J(\br)&=&n_{+\rm b}\left[k_{{\rm s}+}(\br)T_+(\br)-k_{{\rm s}-}(\br)T_-(\br)\right]-q\ce k_{{\rm s}-}(\br)\nonumber\\
&&+q\ce \left(n_{+\rm b}+n_{-\rm b}\right)e^{-V_i(\br)}\int\mathrm{d}\brc k_{\rm sc}(\brc)G(\br,\brc).\nonumber\\
\eea
One notes that the perturbative splitting in Eq.~(\ref{sp}) allowed us to transform the integro-differential Eq.~(\ref{e3}) into the ordinary differential Eqs.~(\ref{e4}) and~(\ref{e6}). Due to their local form, the latter can be easily solved by standard numerical algorithms in order to obtain the net salt-dressed potential~(\ref{sp}).

\subsection{Implementation of the function $T_i(\br)$ in slit pores}
\label{tslit}

We explain here the implementation of the auxiliary function $T_i(\br)$ in Eq.~(\ref{Ts}) in the plane geometry of the nanoslit. To this aim, we split this function into a pore and a bulk part as $T_i(\br)=T_{i{\rm p}}(\br)+T_{\rm ib}(\br)$, where the corresponding components governed by the planar and spherical symmetry, respectively, read
\bea\label{t1}
T_{i{\rm p}}(\br)&=&\int\mathrm{d}\br\ce k_{\rm sc}(\brc)f_i(\br,\brc),\\
\label{t2}
T_{i{\rm b}}(\br)&=&\int\mathrm{d}\br\ce f_{i{\rm b}}(\br-\brc).
\eea
Next, noting that the planar symmetry implies $G(\br,\br\ce)=G(u,z,z\ce)$ and $f_i(\br,\br\ce)=f_i(u,z,z\ce)$, where we introduced the polar distance variable $u=\sqrt{(x-x')^2+(y-y')^2}$, Eq.~(\ref{t1}) becomes in the polar coordinate system
\be\label{t3}
T_{i{\rm p}}(z)=2\pi\int_0^d\mathrm{d}z\ce k_{\rm sc}(z\ce)\int_0^\infty\mathrm{d}uuf_i(u,z,z\ce).
\ee
Translating now the HC constraint $|\br-\br'|>2a_{\rm hc}$ imposed by the HC potential $w(\br-\br')$ into polar coordinates, Eq.~(\ref{t3}) takes the form
\bea\label{t4}
T_{i{\rm p}}(z)&=&2\pi\int_0^d\mathrm{d}z\ce k_{\rm sc}(z\ce)\\
&&\hspace{1mm}\times\int_0^\infty\mathrm{d}uu\left\{e^{-q_iq\ce G(u,z,z\ce)}\theta(u-u_{\rm min})-1\right\},\nonumber
\eea
with the cut-off distance
\be
u_{\rm min}=\sqrt{4a_{\rm hc}^2-(z-z\ce)^2}\;\theta\left(2a_{\rm hc}-|z-z\ce|\right).
\ee
Finally, noting that the spherical symmetry in the bulk implies $f_{i\rm b}(\br-\br')=f_i(v)$, where $v=|\br-\br'|$, and imposing the steric constraint $v>2a_{\rm hc}$, Eq.~(\ref{t2}) becomes 
\be
\label{t5}
T_{i\rm b}=4\pi\int_0^\infty\mathrm{d}vv^2\left\{e^{-q_iq\ce G_{\rm b}(v)}\theta(v-2a_{\rm hc})-1\right\}.
\ee

\section{Identifying the membrane charge reversal conditions from the dressed-ion approach}
\label{drap}

In this appendix, the DH expansion of the SCPB Eq.~(\ref{e3}) is shown to provide a self-consistent route to the \textit{dressed ion} formalism of Ref.~\cite{Podgornik2011}. Within this formalism of analytical tractability, we estimate the necessary conditions for the emergence of the membrane charge reversal by multivalent counterion addition. 

\subsection{Recovering the dressed-ion theory from the DH expansion of the SCPB Eq.}
\label{dri}

\begin{figure}
\includegraphics[width=.7\linewidth]{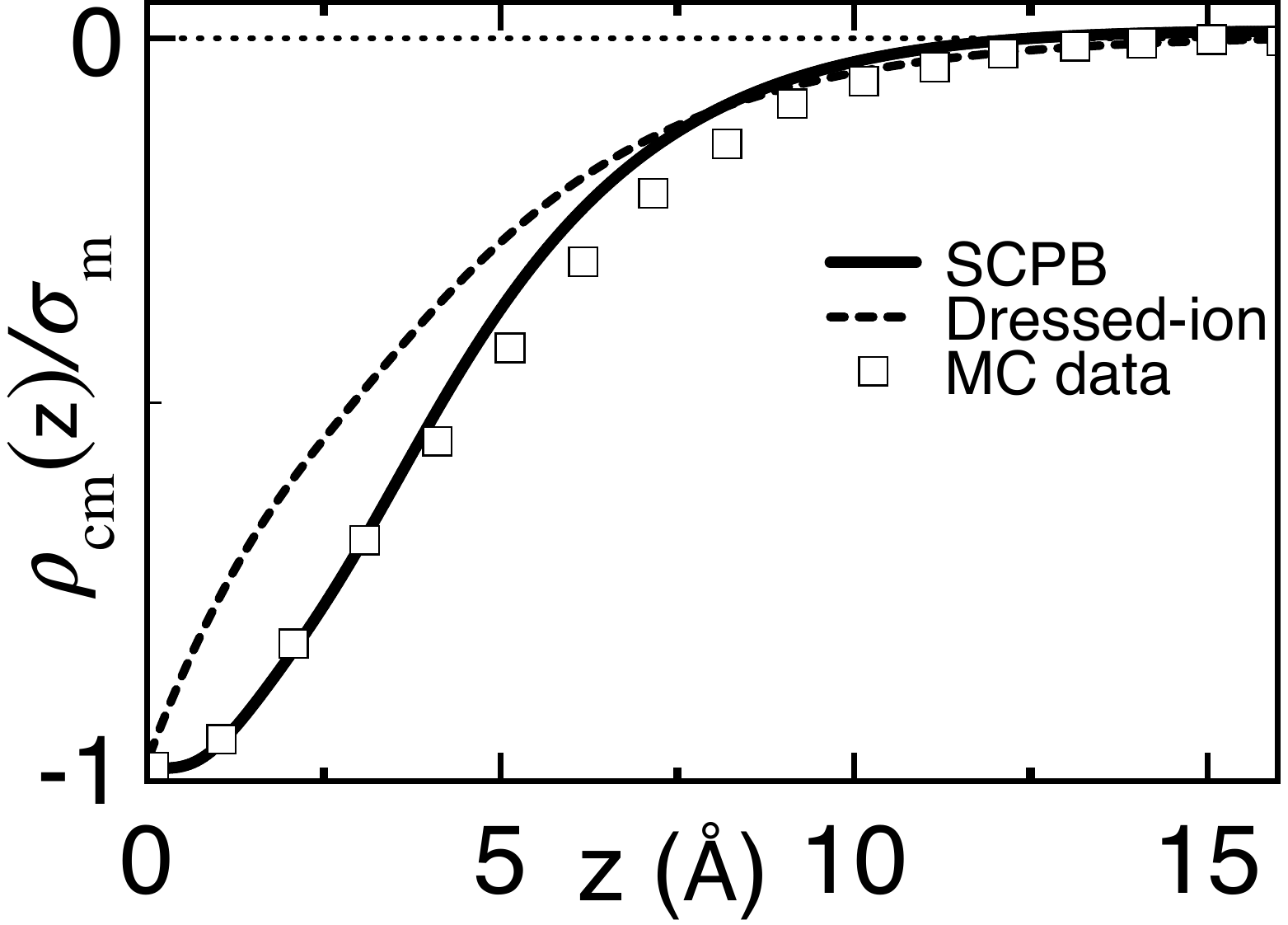}
\caption{(Color online) Cumulative charge density~(\ref{st9}) from the perturbative solution of the SCPB Eq.~(\ref{e3}) explained in Appendix~\ref{per} (solid curve), the dressed ion theory (dashed curve), and the MC simulation data of Ref.~\cite{Podgornik2011}. The model parameters are $\sigma_{\rm m}=0.25$ $e/{\rm nm}^2$, $n_{\rm cb}=0.043$ M, $q\ce=4$, $n_{+b}=0.67$ M, $\e_{\rm m}=0$, $a_{\rm st}=0$, and $a_{\rm hc}=2.25$ {\AA}.}
\label{figAp}
\end{figure}
In order to rederive the dressed-ion formalism, we first use the electroneutrality condition~(\ref{be}) to eliminate the anion concentration from Eq.~(\ref{dh}). This yields the screening parameter $\kappa^2=8\pi\ell_{\rm B}\left(n_{+\rm b}+q\ce n_{\rm cb}/2\right)$ of Ref.~\cite{Podgornik2011}, indicating that the DH Green's Eq.~(\ref{gr1}) and its solution reported in Appendix~\ref{ap1} coincide precisely with their counterpart used in the dressed ion approach.

At the next step, we introduce the MF DH approximation for the monovalent salt ions. This consists in neglecting their self-energy, and also assuming their weak coupling to the charged interfaces, i.e. $|q_\pm\phi\s(\br)|\ll1$. Thus, taking the limit $q_\pm^2\delta G(\br)/2\to 0$, and linearizing the ion densities in Eq.~(\ref{d1}) in terms of the ionic energy $q_\pm\phi\s(\br)$, one obtains
\be\label{dh1}
n_\pm(\br)\approx n_{\pm\rm b}e^{-V_i(\br)}\left[1\mp\phi\s(\br)+n_{\rm cb}T_\pm(\br)\right],
\ee
where we dropped a cross term of order $O\left[q_\pm\phi\s(\br)n_{\rm cb}\right]$. Then, we set the HC interactions to zero, $w(\br-\br')=0$, and Taylor-expand the Mayer function~(\ref{Mayer}) to get $f_i(\br,\brc)\approx-q_iq\ce G(\br,\br')$. Substituting now Eq.~(\ref{dh1}) into the SCPB Eq.~(\ref{e3}), one finds
\bea
\label{dh2}
&&\frac{k_{\rm B} T}{e^2}\nabla_\br\cdot\e(\br)\nabla_\br\phi\s(\br)-\left(n_{+\rm b}+n_{-\rm b}\right)e^{-V_i(\br)}\phi\s(\br)\nonumber\\
&&=-\int\mathrm{d}\br'G^{-1}(\br,\br')\phi\s(\br')=-\sigma(\br),
\eea
where the first equality followed from Eq.~(\ref{grop}). Finally, using Eqs.~(\ref{grop})-(\ref{in}), Eq.~(\ref{dh2}) can be inverted as
\be
\label{dh5}
\phi\s(\br)=\int\mathrm{d}\br'G(\br,\br')\sigma(\br'),
\ee
and the monovalent ion densities in Eq.~(\ref{dh1}) reduce to
\be
\label{dh6}
n_\pm(\br)=n_{\pm\rm b}e^{-V_i(\br)}\left\{\frac{2n_{\mp\rm b}}{n_{+\rm b}+n_{-\rm b}}\mp\Phi(\br)\right\}
\ee
with the net average potential~(\ref{mp2}) taking the form
\be\label{dh7}
\Phi(\br)=\int\mathrm{d}\br'G(\br,\br')\left[\sigma(\br')+q\ce n_{\rm cb}k\ce(\br')\right].
\ee
One can now note that Eqs.~(\ref{dh5})-(\ref{dh7}) correspond precisely to the charge density and the average potential of the dressed ion approach introduced in Ref.~\cite{Podgornik2011}.

Fig.~\ref{figAp} compares the cumulative charge density~(\ref{st9}) obtained from the perturbative solution of the SCPB Eq.~(\ref{e3}) with the prediction of the dressed ion theory and the numerical simulations of Ref.~\cite{Podgornik2011}. One sees that due to the inclusion of the repulsive image charge interactions excluding the ions from the interface, the SCPB theory improves the agreement of the dressed-ion formalism with the simulation data close to the membrane surface. The deviation from the MC data at $z\gtrsim 5$ {\AA} may be due to the incompatibility between the dilute counterion assumption underlying the virial-expanded SCPB approach and the large quadrivalent counterion concentration of the simulations. Additional simulations with dilute counterions will be thus needed to determine the validity regime of the present formalism.

\subsection{Analytical estimation of the membrane charge reversal conditions}

Using here the dressed-ion approach in the regime of large separation distances from the membrane surface, the membrane charge reversal conditions are identified in terms of the surface charge renormalization induced by the multivalent counterions. First, Carrying-out the integral~(\ref{dh5}) with the Green's function~(\ref{Gr}), the salt-dressed potential follows in the single interface limit $d\to\infty$ as the standard DH potential $\phi\s(z)=-2/(\kappa\mu)e^{-\kappa z}$~\cite{Isr}, with the Gouy-Chapman length $\mu=1/(2\pi\ell_{\rm B}\sigma_{\rm m})$~\cite{Gouy,Chapman}. 

For the analytical evaluation of the second term in Eq.~(\ref{dh7}) corresponding to the multivalent ion contribution to the average potential, one has to introduce two additional approximations consistent with the present large distance analysis. First, accounting for the short range of the image-charge and solvation forces~\cite{Buyuk2010}, we neglect the self-energy in Eq.~(\ref{br}) and take $q\ce^2\delta G(\br)/2\to0$. Then, we approximate the DH potential $\phi\s(z)$ by the non-linear solution of the PB Eq. for single interfaces~\cite{Isr}. The counterion partition function~(\ref{br})  becomes $k\ce(z)\approx\left[(1+r)/(1-r)\right]^{2q\ce}$, with the auxiliary variable $r=\gamma e^{-\kappa z}$, and the parameters $\gamma=\sqrt{s^2+1}-s$ and $s=\kappa\mu$. Finally, evaluating the integral in Eq.~(\ref{dh7}) for ${\rm Spm}^{4+}$ cations with $q\ce=4$, in the DH regime $s\gg1$, the net electric field becomes 
\be\label{dhc}
\Phi'(z)\approx2\mu^{-1}e^{-\kappa z}\left(1-\frac{8\pi\ell_{\rm B}q\ce n_{\rm cb}}{\kappa} z\right)+O\left(s^{-1}\right).
\ee

One notes that the second term of Eq.~(\ref{dhc}) corresponding to the multivalent ion contribution to the potential is longer ranged than the pure salt component of opposite sign. This indicates that at large enough separation distances from the membrane surface, any amount of added multivalent cations will result in the reversal of the membrane charge. We emphasize that this peculiarity is in line with our earlier result on the occurrence of like-charge polymer-membrane attraction in the presence of an arbitrary amount of added multivalent counterions under the condition of large enough polymer-substrate separation distance~\cite{Buyuk2019}. Our conclusion based on certain approximations should be of course verified by exact numerical simulations.


\begin{thebibliography}{99}

\bibitem {biomatter} Holm, C.; Kekicheff, P.;  Podgornik, R. {\it Electrostatic Effects in Soft Matter and Biophysics}, Eds.; Kluwer Academic: Dordrecht, 2001.
\bibitem {Schoch} Schoch, R.B.; Han, J.; Renaud, P. {\it Rev. Mod. Phys.} {\bf 2008}, \textit{80}, 839.
\bibitem {Boc1} J.-L. Barrat and  L. Bocquet, \textit{Phys. Rev. Lett.}, 1999, \textbf{82}, 4671-4674.
\bibitem {Boc2} C. Sendner, D. Horinek, L. Bocquet and R. Netz, \textit{Langmuir}, 2009, \textbf{25}, 10768-10781.
\bibitem {Boc3} L. Bocquet and E. Charlaix, \textit{Chem. Soc. Rev.}, 2010, \textbf{39}, 1073-1095.
\bibitem {Boc4} Bocquet, L. {\it Nature Materials} {\bf 2020}, {\it 19}, 254. 
\bibitem {Yar} Yaroshchuk, A. {\it Sep. Purif. Technology} {\bf 2001}, {\it 143}, 22.
\bibitem {Levin2006} Levin, Y. \textit{Europhys. Lett.} \textbf{2006}, \textit{76}, 163.  
\bibitem {Heyden2005} van der Heyden, D. J.; Stein, D.; Dekker, C. {\it Phys. Rev. Lett.} {\bf 2005}, {\it 95}, 116104.  
\bibitem {Daiguji2004} Daiguji, H.; Yang, P.; Szeri, A.J.; Majumdar, A. {\it Nano Lett.} {\bf 2004}, {\it 4}, 2315. 
\bibitem {Heyden2006} van der Heyden, F. H. J.; Bonthuis, D. J.; Stein, D.; Christine, M.; Dekker, C. {\it Nano Lett.} {\bf 2006}, {\it 6}, 2232. 
\bibitem {Heyden2007} van der Heyden, F. H. J.; Bonthuis, D. J.; Stein, D.; Christine, M.; Dekker, C. {\it Nano Lett.} {\bf 2007}, {\it 7}, 1022. 
\bibitem {Gillespie2012} Gillespie, D.  {\it Nano Lett.} {\bf 2012}, {\it 12}, 1410. 
\bibitem {Gillespie2013II} Gillespie, D.; Pennathur, S.; {\it Anal. Chem} {\bf 2013}, {\it 85}, 2991. 
\bibitem {NetzSC} Moreira, A.G.; Netz, R.R. {\it Europhys. Lett.} {\bf 2000}, \textit{52} , 705. 
\bibitem {ExpHey} van der Heyden, F. H. J.; Stein, D.; Besteman, K.; Lemay, S. G.; Dekker, C. \textit{Phys. Rev. Lett.} \textbf{2006}, \textit{96}, 224502.   
\bibitem {e21} Qiu, S.; Wang, Y.; Cao, B.; Guo, Z.; Chen, Y.; Yang, G. \textit{Soft Matter} \textbf{2015}, \textit{11}, 4099. 
\bibitem {Sugimoto2019} Sugimotoa, T.; Nishiyab, M.; Kobayashi, M. {\it Colloids and Surfaces A} {\bf 2019}, {\it 572}, 18. 
\bibitem {Buyuk2018} Buyukdagli, S.  {\it Soft Matter} {\bf 2018}, \textit{14}, 3541. 
\bibitem {Gillespie2011} Gillespie, D.;  Khair, A. S.; Bardhan, J. P. ; and Pennathur, S. \textit{J. Colloid Interface Sci.} {\bf 2011}, {\it 359}, 520. 
\bibitem {Gillespie2013} Hoffmann, J.; Gillespie, D. {\it Langmuir} {\bf 2013}, {\it 29}, 1303. 
\bibitem {Sim1} Gulbrand, L.; J\"{o}nson, B.; Wennerstrom, H.; Linse, B. {\it J. Chem. Phys.} {\bf 1984}, {\bf 82}, 2221. 
\bibitem {Besteman2005} Besteman, K.; Zevenbergen, M. A. G.; Lemay, S.G. {\it Phys. Rev. E} {\bf 2005}, {\it 72}, 061501. 
\bibitem {Forsman2006} Trulsson, M.;  J\"{o}nsson, B.; Torbj\"{o}rn, {\AA}.; Forsman, J.  {\it Phys. Rev. Lett.} {\bf 2006}, {\it 97}, 068302. 
\bibitem {PodWKB} Podgornik, R.; Zeks, B. {\it J. Chem. Soc. Faraday Trans. 2} {\bf 1988}, \textit{84}, 611. 
\bibitem {attard} Attard, P.; Mitchell,  D.J.; Ninham, B.W.; {\it J. Chem. Phys.} {\bf1988}, \textit{89}, 4358. 
\bibitem {netzcoun}  Netz, R.R.; Orland, H.; {\it Eur. Phys. J. E} {\bf 2000}, \textit{1}, 203. 
\bibitem {1loop} Lau, A. W. C. {\it Phys. Rev. E} {\bf 2008}, \textit{77}, 011502. 
\bibitem {Buyuk2012} Buyukdagli, S.; Achim, C.V.; Ala-Nissila {\it J. Chem. Phys.} {\bf 2012}, \textit{137}, 104902. 
\bibitem {Buyuk2014} Buyukdagli, S.; Ala-Nissila, T. \textit{J. Chem. Phys.} \textbf{2014}, \textit{140}, 064701.  
\bibitem {Buyuk2015} Buyukdagli, S.; Blossey, R.; and Ala-Nissila, T. {\it Phys. Rev. Lett.} {\bf 2015}, {\it 114}, 088303. 
\bibitem {Podgornik2010} Kandu\v c, M.; Naji, A.; Forsman, J.; Podgornik, R. {\it J. Chem. Phys.} {\bf 2010} \textit{132}, 124701. 
\bibitem {Podgornik2011} Kandu\v c, M.; Naji, A.; Forsman, J.; Podgornik, R. {\it Phys. Rev. E} {\bf 2011}  \textit{84}, 011502.   
\bibitem {Buyuk2019} Buyukdagli, S.; Podgornik, R. {\it J. Chem. Phys.} {\bf 2019}, \textit{151}, 094902. 
\bibitem {Buyuk2020} Buyukdagli, S.  {\it J. Chem. Phys.} {\bf 2020}, \textit{152}, 014902. 
\bibitem {Storey2012} Storey, B.D.; Bazant, M. Z.; {\it Phys. Rev. E} {\bf 2012}, {\it 86}, 056303. 
\bibitem {Coating} Yuan, J.-K.; Yao, S.-H.; Dang, Z.-M.; Sylvestre, A.; Genestoux, M.; Bai, J. {\it Phys. Chem. C} {\bf 2011}, \textit{115}, 5515. 
\bibitem {NetzHC} Moreira, A.G.; Netz, R.R. {\it Eur. Phys. J. E} {\bf 2002}, \textit{21}, 83.  
\bibitem {Hansen} Hansen, J. P.; McDonald, R. \textit{Theory of Simple Liquids: with Applications to Soft Matter}, Academic Press, 2013. 
\bibitem {rem1} The potential $\phi\s(\br)$ solving the SCPB Eq.~(\ref{e3}) is screened exclusively by the monovalent salt; therein, the multivalent counterions interfere indirectly via their scattering with the monovalent ions.
\bibitem {ionm} The monovalent ion mobilities are set to the experimental values of $\mu_+=7.616\times10^{-8}$ $\mbox{m}^2/\mbox{V}^{-1}\mbox{s}^{-1}$ for ${\rm K}^+$ and $\mu_-=7.909\times10^{-8}$ $\mbox{m}^2\mbox{V}^{-1}\mbox{s}^{-1}$ for $\rm{Cl}^-$~\cite{book}.  To our knowledge, the drift transport mobilities of the phosphate ${\rm PO}^{3-}_4$, spermidine ${\rm Spm}^{3+}$, and spermine ${\rm Spm}^{4+}$ charges are not available in the literature. Based on the Einstein relation $\mu_i=|q_i|D/(k_{\rm B}T)$, the transport coefficient of these charges were estimated as $\mu_{3\pm}=3\mu_{\pm}$  and $\mu_{4+}=4\mu_+$.
\bibitem {book}  Lide, D. R. \textit{Handbook of Chemistry and Physics}, 93th edition, \textbf{2012}, CRC Press.
\bibitem {Butt} Butt, H.-J.; Kappl, M. \textit{Surface and Interfacial Forces}, Wiley-VCH, 2013. 
\bibitem {Arellano2009} Arellano, E. S.-; Olivares, W.; Cassou, M. L.-; Angeles, F. J.- \textit{J. Colloid Interface Sci.} {\bf 2009}, {\it 330}, 474. 
\bibitem {the16} Buyukdagli, S.; Ala-Nissila, T. \textit{Langmuir} \textbf{2014}, \textit{30}, 12907-12915.
\bibitem {rem2} In Appendix~\ref{drap}, a self-consistent rederivation of the dressed-ion theory~\cite{Podgornik2011} from the SCPB Eq.~(\ref{e3}) is carried out. Then, within the dressed-ion approach, it is shown that in contrast with the current reversal phenomenon, CI occurs in the presence of any amount of added multivalent counterions at large enough separation distances from the membrane. This result is an additional argument indicating the absence of one-to-one mapping between CI and current reversal.
\bibitem {Wang2016} Wang, Z.-Y. {\it J. Stat. Mech.} \textbf{2016}, 043205. 
\bibitem {netzvdw} Netz, R.R. {\it Eur. Phys. J. E} {\bf 2001}, \textit{5}, 189. 
\bibitem {Buyuk2011} Buyukdagli, S.; Achim, C.V.; Ala-Nissila,  T. {\it J. Stat. Mech.} \textbf{2011}, P05033. 
\bibitem {Buyuk2010} Buyukdagli, S; Manghi, M.; Palmeri, J. {\it Phys. Rev. E} {\bf 2010}, \textit{81}, 041601. 
\bibitem {Isr} Israelachvili, J. \textit{Intermolecular and Surface Forces}, Academic Press, 1992. 
\bibitem {Gouy} Gouy, G.L. {\it J. Phys.} {\bf 1910} \textit{9}, 457. 
\bibitem {Chapman} Chapman, D.L.; {\it Phil. Mag.} {\bf 1913} \textit{25}, 475. 

\end{thebibliography}
\end{document}